\documentclass[10pt,aps,prd,showpacs,one-column,nofootinbib,noshowkeys,floatfix,preprintnumbers]{revtex4-2}
\usepackage{graphics,graphicx}
\usepackage{amsmath, amssymb}
\usepackage{multirow}
\usepackage{color}
\usepackage{verbatim}
\usepackage{hyperref}
\usepackage[normalem]{ulem}
\usepackage{ulem}
\usepackage{color}
\usepackage{cancel}
\usepackage{mathtools}
\usepackage{epstopdf}
\usepackage{url}
\usepackage{footnote}
\usepackage[absolute]{textpos}
\usepackage{lineno}
\newcommand{\nbcan}{\langle N_B\rangle}
\newcommand{\nabcan}{\langle N_{\bar{B}}\rangle}

\newcommand{\snn}{\ensuremath{ \sqrt{s_{\rm NN} } } }

\newcommand{\Fnorm}[1]{F^{(#1,0)}/F^{(1,0)}}
\newcommand{\Fpr}[1]{F^{(#1,0)}}

\begin{document}

\title{Baselines for Abelian Charge Fluctuations in Nuclear Collisions:\\ Theory and Comparison with Experimental Data}

\author{Bengt Friman}
\affiliation{GSI Helmholtzzentrum f{\"u}r Schwerionenforschung, 64291 Darmstadt, Germany}

\author{Krzysztof Redlich}
\affiliation{Institute of Theoretical Physics, University of Wroc\l{}aw,  50204 Wroc\l{}aw, Poland}

\author{Anar Rustamov}
\affiliation{GSI Helmholtzzentrum f{\"u}r Schwerionenforschung, 64291 Darmstadt, Germany}

\begin{abstract}
We investigate fluctuations in the canonical ensemble of an Abelian charge, such as baryon number. Our focus is on cumulants and factorial cumulants of baryon and antibaryon multiplicity distributions, including their sum and difference, in both the full phase space and subsystems. In particular, we establish a correlation between net-baryon number fluctuations within a subsystem, which is pertinent for fluctuation analyses in nucleus-nucleus collisions, and fluctuations of baryon and antibaryon numbers in the total system. We derive analytical expressions for factorial cumulants of arbitrary order and present concise results in terms of the cumulants of the total baryon number. To account for dynamics beyond global conservation, we introduce local attractive and repulsive multi-particle interactions within a phenomenological framework. A comparison of calculated and generated cumulants with STAR and HADES data indicates that multiparticle interactions play a decisive role in the description of observed fluctuation patterns. At high collision energies, the data are well-reproduced by incorporating repulsive two-proton interactions, while at lower energies, attractive three-particle interactions become essential. Furthermore, our framework facilitates realistic event generation, enabling a direct comparison with experimental measurements.

\end{abstract}
\maketitle 

\section{Introduction}
One of the primary objectives of ongoing experimental and theoretical studies of chiral symmetry restoration in QCD is to map the phase diagram of strongly interacting matter and to determine whether a chiral critical endpoint exists. To address these questions, the STAR collaboration at RHIC launched dedicated research programs, the Beam Energy Scans (BES-I and BES-II), where nucleus-nucleus collisions across a wide range of relativistic energies~\cite{STAR:2021iop, STAR:2020tga, STAR:2025zdq} are explored. Complementary experiments are conducted by several collaborations, HADES at GSI/SIS18~\cite{HADES:2020wpc}, ALICE at CERN/LHC~\cite{ALICE:2019nbs, ALICE:2022xpf} and NA61/SHINE at CERN/SPS~\cite{NA61SHINE:2023gez}, thereby further expanding the energy coverage and augmenting the range of experimental conditions under investigation. 

Fluctuations and correlations of conserved charges have emerged as key observables in this context, offering sensitive probes of the QCD phase structure~\cite{Stephanov:1998dy,Stephanov:1999zu,Asakawa:1989bq,Friman:2011pf,Ejiri:2004bh,Allton:2005gk,Ejiri:2005wq,Karsch:2005ps,Karsch:2010ck,Sasaki:2007db,Sasaki:2006ww,Bzdak:2019pkr,Kuznietsov:2022pcn, Braun-Munzinger:2022bkc, Gross:2022hyw, Rustamov:2020ekv, Rustamov:2022hdi}.
These observables are experimentally accessible and are expected to reflect the critical behavior associated with the chiral phase transition. In particular, fluctuations of the net-baryon number are of great interest, as they are directly linked to critical scaling near the chiral phase boundary~\cite{Friman:2011pf,Karsch:2010ck}. Additional insight into the nature of the chiral phase transition can be obtained by studying multi-particle correlations and the factorial cumulants of baryon and antibaryon multiplicity distributions~\cite{Bzdak:2019pkr,Barej:2020ymr,Bialas:2007ed,Bzdak:2018axe,Barej:2022jij}.
Data on proton, antiproton, and net-proton number fluctuations in heavy-ion collisions have been obtained by the STAR  ~\cite{STAR:2013gus,Luo:2015ewa,Luo:2015doi,STAR:2021iop},  ALICE~\cite{Rustamov:2017lio,Arslandok:2020mda,ALICE:2019nbs}, and HADES~\cite{HADES:2020wpc} Collaborations. These measurements are commonly used as proxies for fluctuations of the baryon, antibaryon, and net-baryon numbers, respectively. The results for the variance, skewness, and kurtosis of the net-proton number are particularly intriguing and have sparked extensive discussion regarding their physical origin and interpretation, with a strong focus on potential signatures of chiral criticality. However, a crucial aspect of the analysis of such fluctuations in heavy-ion collisions is that the measured cumulants are also influenced by effects unrelated to critical phenomena. Two such effects are particularly important. The first is volume fluctuations, which arise from event-by-event variations in the number of participating nucleons~\cite{Skokov:2012ds,Braun-Munzinger:2016yjz,Sugiura:2019toh, Rustamov:2022sqm, Holzmann:2024wyd}. The second involves constraints imposed by exact conservation laws, in particular the conservation of the baryon number in the full phase space~\cite{Braun-Munzinger:2023gsd,Bzdak:2019pkr,Barej:2020ymr,Bialas:2007ed,Bzdak:2018axe,Barej:2022jij,Braun-Munzinger:2016yjz,Bzdak:2012an,Braun-Munzinger:2018yru,Braun-Munzinger:2019yxj,Begun:2004gs,Braun-Munzinger:2020jbk,Vovchenko:2020tsr,Vovchenko:2020gne,Pruneau:2019baa}.

The impact of global conservation laws, particularly the conservation of net-baryon number, on fluctuation observables has been extensively investigated in the context of experimental measurements. These studies examined both cumulants of net-baryon-number fluctuations~\cite{Bzdak:2019pkr,Braun-Munzinger:2016yjz,Bzdak:2012an,Braun-Munzinger:2018yru,Braun-Munzinger:2019yxj,Braun-Munzinger:2020jbk,Vovchenko:2020tsr,Vovchenko:2020gne} and factorial cumulants of baryon and antibaryon multiplicity distributions~\cite{Bzdak:2019pkr,Barej:2020ymr,Barej:2022jij}. It was demonstrated that the observed suppression of net-proton number cumulants, relative to the non-critical (Skellam) baseline, as measured by the STAR BES-I and ALICE experiments in ultra-relativistic nucleus-nucleus collisions, is consistent, within experimental uncertainties, with global conservation of the total net-baryon number~\cite{Braun-Munzinger:2019yxj,Braun-Munzinger:2020jbk}.

However, the situation changes markedly in light of the updated measurements from the STAR BES-II program~\cite{STAR:2025zdq, Zachary:2025qm, Huang:2025qm} and the HADES experiment~\cite{Nabroth:2025qm}. As we demonstrate in this study, the non-critical baseline expectations, including those incorporating global baryon number conservation, fail to reproduce the experimental results for proton factorial cumulants. In fact, ratios of factorial cumulants exhibit substantial deviations from the corresponding baselines based on global conservation laws. At higher collision energies, such discrepancies have already been reported. A significantly improved description of the data was achieved by accounting for the finite hadron volumes within the statistical approach~\cite{Vovchenko:2021kxx}.

In this study, we extend previous research on the impact of baryon number conservation on fluctuation observables. We establish a link between fluctuations of the net-baryon number within a subsystem, pertinent to fluctuation studies in nucleus-nucleus collisions, and fluctuations of the baryon and antibaryon numbers in the entire system. Furthermore, considering exact baryon number conservation, we compute factorial cumulants of the baryon, antibaryon, mixed baryon–antibaryon, and net-baryon number multiplicity distributions within a subsystem. General expressions for these factorial cumulants are derived to arbitrary order, resulting in concise results expressed in terms of the cumulants of the total-system baryon number. These findings complement the earlier work by Barej and Bzdak~\cite{Barej:2020ymr,Barej:2022jij}, where analytical results were obtained up to the sixth order.

One of the novel aspects of this study is the simultaneous inclusion of both repulsive and attractive interactions among baryons within the canonical ensemble -- a feature not addressed in previous analyses. Furthermore, we incorporate interactions involving multiple baryons, enabling the exploration of baryon correlations, which is crucial for experimental searches for phase transitions. We demonstrate that, with these extensions, the energy dependence of proton number factorial cumulants observed in the STAR BES-II program, as well as in the HADES results, can be understood as the result of an interplay between attractive and repulsive interactions among protons, implemented consistently within the canonical ensemble. We note that attractive interactions between baryon pairs in the canonical ensemble were recently investigated in Ref.~\cite{Braun-Munzinger:2023gsd}. In this study, we extend this approach further to account for repulsive interactions as well as interactions involving multi-particle states.

This manuscript substantially expands the study of canonical effects on fluctuations, presented in~\cite{Friman:2022wuc}, by incorporating novel advancements, particularly the inclusion of correlations generated by local attractive and repulsive multiparticle interactions. Our findings indicate that such correlations are crucial for obtaining a quantitative description of fluctuation observables.  

The paper is organized as follows: In the subsequent section, we introduce the canonical partition function for uncorrelated baryons with exact net-baryon number conservation and discuss the relevant fluctuation observables. In Section~\ref{Cumulant_generating_functions}, we formulate the cumulant and factorial cumulant generating functions. Analytic expressions for factorial cumulants within a subsystem are derived in Section~\ref{Fluctuations_in_a_subsystem}. In Section~\ref{Local_conservation_laws}, we introduce repulsive and attractive correlations among baryons. A comparison with experiment is presented in Section~\ref{Comparison_with_experimental_results}. Finally, we summarize our findings in Section~\ref{Summary}. Mathematical details and general results independent of the specific form of the partition function are given in three appendices.

%%%%%%%%%%%%%%%%%%%%%%%%%%%%%%%%%%%%%%%%%%%%%%%%%%%%%%%%

\section{Canonical ensemble of the net-baryon number}
\label{sec_canonical}
In order to obtain analytical results on the influence of exact charge conservation on fluctuation observables, we adopt a thermal model for the net-baryon number conservation in the Boltzmann approximation and neglect baryon-baryon interactions\footnote{The effect of interactions between baryons is addressed in Section~\ref{Local_conservation_laws}.}.  The statistical operator is formulated following 
the S-matrix approach \cite{Dashen:1969ep,Venugopalan:1992hy,Weinhold:1997ig,Giacosa:2016rjk,Lo:2017sde,Dash:2018mep,Friman:2015zua, Lo:2020phg}, where to leading order in the fugacity expansion it has the form of an ideal gas, albeit with  the thermal phase space of free baryons modified by meson-baryon interactions. 

The S-matrix thermodynamic potential 
reproduces particle production yields in heavy-ion collisions  \cite{ Braun-Munzinger:2003pwq,Andronic:2017pug,Cleymans:2020fsc, Andronic:2018qqt}, and %as well as
describes basic properties of the net proton number cumulants and their energy dependence, as obtained by the STAR collaboration during its BES-I program \cite{ Braun-Munzinger:2020jbk} and by the ALICE collaboration.  Moreover, the S-matrix thermodynamic potential of a hadron gas formulated in the grand canonical ensemble is consistent with the lattice QCD  equation of state and certain second-order cumulants and correlations of conserved charges in the confined phase \cite{Bazavov:2017dus, Noronha-Hostler:2019ayj, Lo:2017lym,Almasi:2019yaw,Goswami:2020yez}.

In the grand canonical partition function
\begin{eqnarray}\label{eq:grand-canonical-partition}
    \mathcal{Z}(\mu_B,T)&=&\sum_{N_{B}=0}^{\infty}\sum_{N_{\bar{B}}=0}^{\infty}\frac{(e^{\mu_B/T}\,z_B)^{N_{B}}}{N_{B}!}\frac{(e^{-\mu_B/T}\,z_{\bar{B}})^{N_{\bar{B}}}}{N_{\bar{B}}!} \nonumber\\
    &=&\exp\big(e^{\mu_B/T}\,z_B+e^{-\mu_B/T}\,z_{\bar B}\big).
\end{eqnarray}
the fluctuations of baryons and antibaryons are described by two independent Poisson distributions. 
In the spirit of the S-matrix approach, the effect of meson-baryon interactions is subsumed in the baryon and antibaryon single-particle partition functions $z_B$ and $z_{\bar B}$. 

The canonical partition function in the ensemble, where the net-baryon number is conserved, is given by \cite{Braun-Munzinger:2020jbk,Braun-Munzinger:2003pwq} \footnote{As noted in~\cite{Braun-Munzinger:2020jbk}, this form of the partition function applies also to systems where the composition is non-uniform, provided the baryon and antibaryon multiplicities are locally Poisson distributed. Consequently, it is more general than the global thermal statistical model employed as a motivation.}
\begin{eqnarray}%\label{eq:canonical-partition}
    Z_{B}&=&\sum_{N_{B}=0}^{\infty}\sum_{N_{\bar{B}}=0}^{\infty}\frac{(\lambda_{B}\,z_B)^{N_{B}}}{N_{B}!}\frac{(\lambda_{\bar{B}}\,z_{\bar{B}})^{N_{\bar{B}}}}{N_{\bar{B}}!}\delta({N_{B}-N_{\bar{B}}-B}) \nonumber\\
    &=&\int_0^{2\pi}\frac{d\phi}{2\pi}\,e^{-iB\phi}\exp\left({\lambda_B\,z_B\,e^{i\phi}+\lambda_{\bar B}\,z_{\bar{B}}\,e^{-i\phi}}\right)
    \label{eq:canonical-partition}\\
    &=&\left(\frac{\lambda_{B}\,z_B}{\lambda_{\bar{B}}\,z_{\bar{B}}}\right)^{\frac{B}{2}}\,I_{B}(2\,z\,\sqrt{\lambda_{B}\,\lambda_{\bar{B}}}),\nonumber
\end{eqnarray}
where the auxiliary parameters $\lambda_{B,\bar{B}}$ are introduced for the calculation of the mean number of baryons and antibaryons and the corresponding cumulants,
\begin{eqnarray}\label{NCM}
    \langle N_B\rangle&=& \lambda_B{\frac{\partial\ln Z_B} {\partial\lambda_B}}|_{\lambda_B,\lambda_{\bar B}=1}=z\,\frac{I_{B-1}(2\, z)}{I_{B}(2\, z)},\\\label{NCP}
    \langle N_{\bar{B}}\rangle&=&\lambda_{\bar B}{\frac{\partial\ln Z_B} {\partial\lambda_{\bar B}}}|_{\lambda_B,\lambda_{\bar B}=1}=z\,\frac{I_{B+1}(2\, z)}{I_{B}(2\, z)},\\
    %\end{eqnarray}
    %\begin{eqnarray}
    \langle (\delta N_{\rm B})^k\rangle_c&=&\left[\left(\lambda_B\,\frac{\partial}{\partial  \lambda_B}\right)^{k}\log Z_B\right]_{\lambda_B,\lambda_{\bar B}=1}\qquad (k>1).\label{eq:cumulant-lambda-deriv}
\end{eqnarray}
Here (and below) $\langle \dots \rangle$ denotes an expectation value in the canonical ensemble, while $\langle \dots \rangle_c$ is the connected part thereof. Thus, fluctuations of the baryon and antibaryon numbers in the canonical ensemble are given by $\delta N_B=N_B-\langle N_B\rangle$ and $\delta N_{\bar B}=N_{\bar B}-\langle N_{\bar B}\rangle$,
and (\ref{eq:cumulant-lambda-deriv}) yields the cumulants of the number of baryons in the full system.

We can rewrite (\ref{eq:cumulant-lambda-deriv}) in terms of the mean baryon and antibaryon multipicities, keeping the dependence on the auxiliary parameters $\lambda_{\rm B},\lambda_{\bar{\rm B}}$
\begin{equation}\label{eq:cumulant-recurrence}
    \langle (\delta N_{\rm B})^k\rangle_c=\frac12\,\left[\left(\lambda_B\,\frac{\partial}{\partial  \lambda_B}\right)^{k-1}\left(\langle N_B\rangle_\lambda+\langle N_{\bar{B}}\rangle_\lambda\right)\right]_{\lambda_B,\lambda_{\bar B}=1}\quad (k>1),
\end{equation}
where
\begin{eqnarray}\label{NCM-2}
    \langle N_B\rangle_\lambda&=&z\,\sqrt{\lambda_B\,\lambda_{\bar B}}\,\frac{I_{B-1}(2\, z\,\sqrt{\lambda_B\,\lambda_{\bar B}})}{I_{B}(2\, z\,\sqrt{\lambda_B\,\lambda_{\bar B}})},\\\label{NCP-2}
    \langle N_{\bar{B}}\rangle_\lambda&=&z\,\sqrt{\lambda_B\,\lambda_{\bar B}}\,\frac{I_{B+1}(2\, z\,\sqrt{\lambda_B\,\lambda_{\bar B}})}{I_{B}(2\, z\,\sqrt{\lambda_B\,\lambda_{\bar B}})}.
\end{eqnarray}
Hence, the baryon and antibaryon multiplicities, (\ref{NCM-2}) and (\ref{NCP-2}), are functions of $z\sqrt{\lambda_B\,\lambda_{\bar B}}$. This reflects the fact that in the canonical ensemble $N_B-N_{\bar B}$ is conserved and thus does not fluctuate~\footnote{This is easily seen by noting that the recurrence relation $2\,\nu\, I_\nu(x)=x\,\big[I_{\nu-1}(x)-I_{\nu+1}(x)\big]$ implies that $\langle N_B\rangle_\lambda-\langle N_{\bar{B}}\rangle_\lambda=B$ is independent of $\lambda_B$ and $\lambda_{\bar B}$.}. Consequently, the fluctuations of $N_B$ and $N_{\bar B}$ are equal, i.e., $\delta N_{\rm B}=\delta N_{\bar{\rm B}}$ and cumulants of the form
\begin{equation}\label{eq:canonical-cumulant-eq}
    \langle (\delta N_{\rm B})^{n-m}\,(\delta N_{\bar{\rm B}})^m\rangle_c
\end{equation}
are independent of $m$, and thus all equal to $\langle (\delta N_{\rm B})^{n}\rangle_c$. In the {\em full} system, it is therefore sufficient to consider only cumulants of the number of baryons. 

Moreover, it follows that the derivatives of $\langle N_B\rangle_\lambda+\langle N_{\bar{B}}\rangle_\lambda$ with respect to $\lambda_B$ can be replaced by derivatives with respect to $z$, setting $\lambda_B=\lambda_{\bar B}=1$. 
Using the notation
\begin{align}\label{eq:shorthand-1}
    c_1=&\frac12 \langle N_{\rm B} + N_{\bar{\rm B}}\rangle,\\
    c_k=&\langle (\delta N_{\rm B})^k\rangle_c\qquad (k>1), \label{eq:shorthand-2}
\end{align}
we can then rewrite (\ref{eq:cumulant-recurrence}) as
\begin{equation}\label{eq:cumulants-NB}
    c_k=\left(\frac{z}{2}\,\frac{d}{d  z}\right)^{k-1}c_1\qquad (k>1).
\end{equation}
Clearly, the cumulants satisfy the recurrence relation
\begin{equation}\label{eq:rec-rel-n}
    c_{k+1}=\frac{z}{2}\,\frac{d}{d z}\, c_k.
\end{equation}

It is useful to define $N^{(tot)}=N_B+N_{\bar B}$ and to consider the  fluctuations thereof, $\delta N^{(tot)}=\delta N_B + \delta N_{\bar B}$. We introduce the notation
\begin{eqnarray}\label{eq:cumulants-Ntot}
    C_1&=&\langle N^{(tot)}\rangle,\\
    C_k&=&\langle (\delta N^{(tot)})^k\rangle_c \qquad (k>1). \nonumber
\end{eqnarray}
Using the fact that $\delta N_{\rm B}=\delta N_{\bar{\rm B}}$ and (\ref{eq:shorthand-1},\ref{eq:shorthand-2},\ref{eq:rec-rel-n}), it follows that
\begin{equation} \label{eq:relation-cumulants}
    C_k=2^k\,c_k
\end{equation}
and
\begin{equation}\label{eq:rec-rel-N}
    C_{k+1}=z\,\frac{d}{d z}\, C_k.
\end{equation}

It is straightforward to compute the baryon number cumulants $c_k$ using  the recurrence relation for $c_k$, (\ref{eq:rec-rel-n}). Given that $c_1=\frac12(N_B+N_{\bar B})$, and using the notation~\cite{Braun-Munzinger:2020jbk}
\begin{eqnarray}\label{eq:SPQW}
    S&=&\langle N_B+N_{\bar B}\rangle,\nonumber\\
    P&=&\langle N_B\rangle\,\langle N_{\bar B}\rangle,\\
    Q&=&z^2-P,\nonumber\\
    W&=&Q\,S-P.\nonumber
\end{eqnarray}
with the derivatives
\begin{equation}\label{eq:SPQW-deriv}
    S'=\frac{4}{z}\,Q,\qquad  P'=\frac{2}{z}\,Q\,S,\qquad Q'=\frac{2}{z}\,\big(Q-W\big),
\end{equation}
one finds for the first few
\begin{eqnarray}\label{eq:n_S_Q-W}
    c_1&=&S/2,\nonumber\\
    c_2&=&Q,\\
    c_3&=&Q-W,\nonumber\\
    c_4&=&Q-W\,+W\,S-2Q^2.\nonumber
\end{eqnarray}
The corresponding expressions for the cumulants $C_k$ are trivially obtained using (\ref{eq:relation-cumulants}).

\section{Cumulant generating functions}\label{sect:cum-gen-functs}
\label{Cumulant_generating_functions}
The generating function for the cumulants of $N^{(tot)}=N_B+N_{\bar B}$, (\ref{eq:cumulants-Ntot}),  in the canonical ensemble with net-baryon number $B$ is obtained by evaluating
\begin{eqnarray}\label{eq:can_cumulant_gen_function}
    G_c(t)&=&\ln\Big[\sum_{N_{B}=0}^{\infty}\sum_{N_{\bar{B}}=0}^{\infty}\frac{(z_B)^{N_{B}}}{N_{B}!}\frac{(z_{\bar{B}})^{N_{\bar{B}}}}{N_{\bar{B}}!}\delta({N_{B}-N_{\bar{B}}-B})\,e^{(N_B+N_{\bar B})\,t}\Big]\nonumber\\
    &=&\ln\Big[\int_0^{2 \pi}\frac{d \phi}{2\,\pi}\,\exp\big(z_B\,e^{t+i \phi}\big)\,\exp\big(z_{\bar B}\,e^{t-i \phi}\big)\,e^{-i\,\phi\,B}\Big]\\
    &=&\ln\Big[\left(\frac{z_B}{z_{\bar B}}\right)^{B/2}\,I_B\big(2\,z\,e^t\big)\Big],\nonumber
\end{eqnarray}
where we set the auxiliary parameters $\lambda_B,\lambda_{\bar B}$ to unity. The $k$:th cumulant is then given by
\begin{equation}\label{eq:can-cumulant-Ntot}
    C_k=\frac{d^k G_c(t)}{d\,t^k}\mid_{t=0}.
\end{equation}

The generating functions for the corresponding factorial cumulants is obtained by the replacement~\cite{Kitazawa:2017ljq} $e^t\to x$,
\begin{equation}
    H_c(x)=\ln\Big[\left(\frac{z_B}{z_{\bar B}}\right)^{B/2}\,I_B\big(2\,z\,x\big)\Big].
\end{equation}
The factorial cumulants of $\delta(N_B+N_{\bar B})$ are then given by
\begin{equation}\label{eq:factorial-cumulants-tot}
    F_n=\frac{d^n H_c(x)}{d\,x^n}\mid_{x=1}.
\end{equation}
We note that the factorial cumulants are identical to the functions $f^{(n)}(z)$ defined in \cite{Braun-Munzinger:2020jbk} and thus satisfy the recurrence relation
\begin{equation}\label{eq:rec-relation-F}
    F_{n+1}=z^{n+1}\frac{d}{d\,z}\left(F_{n}/z^n\right).
\end{equation}
The factorial cumulant $F_n$ of any order $n$ can be obtained by using (\ref{eq:rec-relation-F}) and (\ref{eq:SPQW},\ref{eq:SPQW-deriv}), starting from $F_1=C_1=S$.

The cumulants and factorial cumulants in the full system are connected via the general relations \begin{equation}\label{eq:Cn_to_Nk-relation}
    F_n=\sum_{k=1}^n\,s(n,k)\,C_k,
\end{equation}
and
\begin{equation}\label{eq:Nn-Ck-relation}
    C_n=\sum_{k=1}^n\,S(n,k)\,F_k,
\end{equation}
derived in~\ref{sect:App-c}, where $s(n,k)$ and $S(n,k)$ are Stirling numbers of the first and second kind, respectively. 

\section{Fluctuations in a subsystem}
\label{Fluctuations_in_a_subsystem}
In this section, we explore the fluctuations in a subsystem, $A$, of the canonical system discussed so far. The part of the system that is not included in $A$ we denote by $R$. While the net-baryon number is conserved in the full system, $A+R$, it is not in the subsystem because net-baryon number can be transferred between $A$ and $R$. We obtain analytic results for the cumulants and factorial cumulants of the baryon and antibaryon numbers in $A$. Furthermore, we establish general relations between these cumulants and those of the net-baryon number. 

The cumulants of fluctuations within a subsystem are computed by differentiating the corresponding generating functions. Thereby, we employ the Fa\`{a} di Bruno formula~\cite{Riordan:1946,Comtet:1974}
\begin{equation}\label{eq:faa-di-bruno-1}
    \frac{d^n}{dx^n}f(h(x))=\sum_{k=1}^n f^{(k)}(h(x))\,B_{n,k}\left(h^{(1)}(x),h^{(2)}(x),\dots,h^{(n-k+1)}(x)\right),
\end{equation}
and generalizations thereof~\cite{Riordan:1946,Schumann:2019xy}.
In (\ref{eq:faa-di-bruno-1}), $f^{(k)}$ and $h^{(k)}$ denote the $k$th derivatives, while $B_{n,k}(y_1,y_2,\dots)$ are partial Bell polynomials~\cite{Bell:1934,Comtet:1974}.

\subsection{Factorial cumulants of \texorpdfstring{$\delta N_B$}{d N B} and \texorpdfstring{$\delta N_{\bar B}$}{d N B}}\label{sec:fac-cum}

Analytic expressions for the factorial cumulants of the baryon and antibaryon numbers up to sixth order in a subsystem, taking the exact conservation of the net-baryon number in the total system into account, were presented in~\cite{Barej:2020ymr}. In this section, we derive closed-form expressions for these factorial cumulants. Compact expressions are obtained in  terms of the cumulants of the baryon number of the total system, $c_k$.

The generating function for factorial cumulants is in this case~\cite{Barej:2020ymr}
\begin{eqnarray}\label{eq:generating-function-fact-cum}
    g_{fc}(x,\bar{x})&=&\frac{B}{2}\big[\ln\big(p(x)\big)-\ln\big(\bar{p}(\bar{x})\big)\big]\\
    &+&\ln\big[I_B\big(2\,z\,\sqrt{p(x)\,\bar{p}(\bar{x})}\big)\big],\nonumber
\end{eqnarray}
where $B$ is the total net-baryon number, $p(x)=1-\alpha_B+\alpha_B\,x$ and $\bar{p}(\bar{x})=1-\alpha_{\bar B}+\alpha_{\bar B}\,\bar{x}$. Here  $\alpha_B$ is the probability for finding a baryon in the subsystem $A$ and $\alpha_{\bar B}$ that for an antibaryon, while $z=\sqrt{z_B\,z_{\bar{B}}}$ is the geometric mean of the single-particle partition functions $z_B$ and $z_{\bar{B}}$ (cf. Ref.~\cite{Braun-Munzinger:2020jbk}).

The factorial cumulants within the subsystem $A$ are computed by evaluating the derivatives with respect to $x$ and $\bar{x}$
\begin{equation}\label{eq:factorial-cumulants}
    F^{(n,m)}=\frac{\partial^n}{\partial\,x^n}\,\frac{\partial^m}{\partial \bar{x}^m}\,g_{fc}(x,\bar{x})|_{x=\bar{x}=1}.
\end{equation}

The derivatives of the first two terms in (\ref{eq:generating-function-fact-cum}), $g_{fc,1}(x)$ and $g_{fc,2}(\bar{x})$, are obtained by employing the Fa\`{a} di Bruno formula for functions of the form $\log(f(y))$, where $f(y)$ is a first-order polynomial in $y$~\cite{Comtet:1974}. One finds
\begin{equation}
    \frac{\partial^n}{\partial x^n}\,g_{fc,1}(x)=-\frac{B}{2}\,(-\alpha_B)^n\,(n-1)!,
\end{equation}
and
\begin{equation}
    \frac{\partial^n}{\partial \bar{x}^n}\,g_{fc,2}(\bar{x})=\frac{B}{2}\,(-\alpha_{\bar B})^n\,(n-1)! .
\end{equation}

For the last term in (\ref{eq:generating-function-fact-cum}),
\begin{equation}\label{eq:fc-gen-func-c-term}
    g_{fc,3}(x,\bar{x})=\ln\big[I_B\big(2\,z\,\sqrt{p(x)\,\bar{p}(\bar{x})}\big)\big],
\end{equation}
we consider the composite function $f_3( g_3(x,\bar{x}))$, where $f_3( y)=\ln\big[I_B(2\,z\,e^y)\big]$ and $g_3(x,\bar{x})=\frac12\,\big(\ln\big[p(x)\big]+\ln\big[\bar{p}(\bar{x})\big]\big)$. The derivatives of $f_3(y)$ are obtained using (\ref{eq:can_cumulant_gen_function}) and (\ref{eq:can-cumulant-Ntot}), which yield $f^{(n)}_3(0)= C_n$. Furthermore, the non-zero derivatives of $g_3(x,\bar{x})$ are given by
\begin{eqnarray}\label{eq:gc-derivatives}
    g_3^{(n,0)}&=&-\frac12(-\alpha_B)^n\,(n-1)!\quad (n>0)\\
    g_3^{(0,m)}&=&-\frac12(-\alpha_{\bar B})^m\,(m-1)!\quad (m>0).\nonumber
\end{eqnarray}
Now, using the multivariate Fa\`{a} di Bruno formula to compute the derivatives with respect to $x$ and $\bar{x}$ of $g_{fc,3}(x,\bar{x})$ we find
\begin{equation}
    \frac{\partial^n}{\partial\,x^n}\,\frac{\partial^m}{\partial \bar{x}^m}\,g_{fc,3}(x,\bar{x})=\sum_{k=1}^{n+m}\,C_k\,B_{n,m;k}(\{g_3^{(i,j)}\})
\end{equation}
where the $B_{n,m;k}(\{g_3^{(i,j)}\})$ are multivariate Bell polynomials (see appendix \ref{sec:app-a} and Ref.~\cite{Schumann:2019xy}). As shown in appendix \ref{sec:app-b}, the multivariate Bell polynomials are, for the derivatives (\ref{eq:gc-derivatives}), given by
\begin{equation}
    B_{n,m;k}(\{g_3^{(i,j)}\})=\frac{(\alpha_B)^n\,(\alpha_{\bar B})^m}{2^k}\,\,s(n,m;k),
\end{equation}
where $s(n,m;k)$ are generalized Stirling numbers of the first kind. They 
can be expressed in terms of the standard Stirling numbers of the first kind,
\begin{equation}
    s(n,m;k)=\sum_{l=0}^{k}\,s(n,k-l)\,s(m,l).
\end{equation}

Now, collecting all terms and using (\ref{eq:relation-cumulants}), we find a closed-form expression for the factorial cumulants of a subsystem, which accounts for baryon-number conservation in full phase space,
\begin{eqnarray}\label{eq:factorial-cumulants-closed-form}
    F^{(n,m)}&=&-\frac{B}{2}\Big((-\alpha_B)^{n}\,(n-1)!\,\delta_{m,0}-(-\alpha_{\bar B})^{m}\,(m-1)!\,\delta_{n,0}\Big)\nonumber\\
    &+&(\alpha_B)^n\,(\alpha_{\bar B})^m\sum_{k=1}^{n+m}c_k\sum_{l=0}^{k}\,s(n,k-l)\,s(m,l).
\end{eqnarray}
The first few factorial cumulants are
\begin{eqnarray}\label{eq:factorial-cumulants-3}
    F^{(1,0)}&=&\alpha_B\,\nbcan,\nonumber\\
    F^{(0,1)}&=&\alpha_{\bar B}\,\nabcan,\nonumber\\
    F^{(2,0)}&=&(\alpha_B)^2\,\big(c_2-\nbcan\big),\nonumber\\
    F^{(1,1)}&=&\alpha_B\,\alpha_{\bar B}\,c_2,\nonumber\\
    F^{(0,2)}&=&(\alpha_{\bar B})^2\,\big(c_2-\nabcan\big),\\
    F^{(3,0)}&=&(\alpha_B)^3\,\big(c_3-3\,c_2+2\,\nbcan\big),\nonumber\\
    F^{(2,1)}&=&(\alpha_B)^2\,\alpha_{\bar B}\,\big(c_3-c_2\big),\nonumber\\
    F^{(1,2)}&=&\alpha_B\,(\alpha_{\bar B})^2\,\big(c_3-c_2\big),\nonumber\\
    F^{(0,3)}&=&(\alpha_{\bar B})^3\,\big(c_3-3\,c_2+2\,\nabcan\big).\nonumber
\end{eqnarray}\\

\subsection{Cumulants of \texorpdfstring{$\delta N_B$}{d N B} and \texorpdfstring{$\delta N_{\bar B}$}{d N B}}

The generating function for cumulants of $\delta N_B$ and $\delta N_{\bar B}$ in the subsystem is obtained from (\ref{eq:generating-function-fact-cum}) by the replacements $x\to e^t$ and $\bar{x}\to e^s$,
\begin{eqnarray}\label{eq:generating-function-NB-cum}
    g_{c}(t,s)&=&\frac{B}{2}\big[\ln\big(q(t)\big)-\ln\big(\bar{q}(s)\big)\big]\\
    &+&\ln\big[I_B\big(2\,z\,\sqrt{q(t)\,\bar{q}(s)}\big)\big],\nonumber
\end{eqnarray}
where $q(t)=1-\alpha_B+\alpha_B\,e^t$ and $\bar{q}(s)=1-\alpha_{\bar B}+\alpha_{\bar B}\,e^s$. 

The cumulants in the subsystem $A$, defined by the acceptance probabilities $\alpha_B$ and $\alpha_{\bar B}$,
are given by
\begin{equation}
    C^{(n,m)}=\frac{\partial^n}{\partial t^n}\,\frac{\partial^m}{\partial s^m}\,g_c(t,s)|_{t=s=0}.
\end{equation}

As shown in appendix \ref{sect:App-c}, the relation between the generating functions, $g_c(t,s)=g_{fc}(e^t,e^s)$, implies that the cumulants can be obtained from the corresponding factorial cumulants using
\begin{equation}\label{eq:cumulant-fac-cumulant-relation-subsystem-text}
    C^{(n,m)}=\sum_{k_1=0}^n\,\sum_{k_2=0}^m\,F^{(k1,k2)}\,S(n,k_1)\,S(m,k_2).
\end{equation}
By substituting the expression for the factorial cumulants in the canonical ensemble (\ref{eq:factorial-cumulants-closed-form}) into (\ref{eq:cumulant-fac-cumulant-relation-subsystem-text}), we obtain
\begin{eqnarray}\label{eq:baryon-antib-cumulants-sub}
    C^{(n,m)}&=&\frac{B}{2}\,\Big(\kappa_{\rm Ber}^{(n)}(\alpha_B)\,\delta_{m,0}-\kappa_{\rm Ber}^{(m)}(\alpha_{\bar B})\,\delta_{n,0}\Big)\\
    &+&\sum_{k=1}^{n+m}c_k\sum_{i=0}^k B_{n,k-i}\big(\kappa_{\rm Ber}^{(1)}(\alpha_B),\kappa_{\rm Ber}^{(2)}(\alpha_B),\dots\big)\nonumber\\
    &\times&B_{m,i}\big(\kappa_{\rm Ber}^{(1)}(\alpha_{\bar B}),\kappa_{\rm Ber}^{(2)}(\alpha_{\bar B}),\dots\big),\nonumber
\end{eqnarray}
where \cite{Braun-Munzinger:2020jbk}
\begin{equation}\label{eq:Bernoulli-cumulant}
    \kappa_{\rm Ber}^{(n)}(\alpha)=\delta_{n,1}+(-1)^{1+n}\,{\rm Li}_{1-n}(1-1/\alpha),
\end{equation}
are the cumulants of the Bernoulli distribution with success probability $\alpha$ and ${\rm Li}_{n}(x)$ is the polylogarithm. 
In deriving (\ref{eq:baryon-antib-cumulants-sub}) we employed the relations
\begin{eqnarray}
    &&\sum_{k=1}^n S(n,k)\,(-\alpha)^k\,(k-1)!=-\kappa^{(n)}_{\rm Ber}(\alpha),\\
    &&\sum_{k=1}^n S(n,k)\,(\alpha)^k\,s(k,i)=B_{n,i}\big(\kappa_{\rm Ber}^{(1)}(\alpha),\kappa_{\rm Ber}^{(2)}(\alpha),\dots\big).\nonumber
\end{eqnarray}
We provide explicit expressions for the initial few cumulants in the subsystem,
\begin{eqnarray}\label{eq:cumulants-3}
    C^{(1,0)}&=&\alpha_B\,\nbcan,\nonumber\\
    C^{(0,1)}&=&\alpha_{\bar B}\,\nabcan,\nonumber\\
    C^{(2,0)}&=&(\alpha_B)^2\,c_2+\alpha_B(1-\alpha_B)\,\nbcan,\nonumber\\
    C^{(1,1)}&=&\alpha_B\,\alpha_{\bar B}\,c_2,\nonumber\\
    C^{(0,2)}&=&(\alpha_{\bar B})^2\,c_2+\alpha_{\bar B}(1-\alpha_{\bar B})\,\nabcan,\\
    C^{(3,0)}&=&(\alpha_B)^3\,c_3+3(\alpha_B)^2\,(1-\alpha_B)\,c_2\nonumber\\
    &+&\alpha_B(1-3\,\alpha_B+2\,(\alpha_B)^2)\,\nbcan,\nonumber\\
    C^{(2,1)}&=&(\alpha_B)^2\,\alpha_{\bar B}\,c_3+\alpha_B\,\alpha_{\bar B}\,(1-\alpha_B)\,c_2,\nonumber\\
    C^{(1,2)}&=&\alpha_B\,(\alpha_{\bar B})^2\,c_3+\alpha_B\,\alpha_{\bar B}\,(1-\alpha_{\bar B})\,c_2,\nonumber\\
    C^{(0,3)}&=&(\alpha_{\bar B})^3\,c_3+3(\alpha_{\bar B})^2\,(1-\alpha_{\bar B})\,c_2\nonumber\\
    &+&\alpha_{\bar B}(1-3\,\alpha_{\bar B}+2\,(\alpha_{\bar B})^2)\,\nabcan.\nonumber
\end{eqnarray}

\subsection{Cumulants of the net-baryon number}

The generating function for net-baryon number cumulants  within a subsystem is obtained by making the substitutions $x\to e^t$ and $\bar{x}\to e^{-t}$ in (\ref{eq:generating-function-fact-cum}). Alternatively, the generating function can be derived from the probability distribution $P_A(B_A)$ of the net-baryon number within the subsystem~\cite{Bzdak:2012an,Braun-Munzinger:2020jbk}. One finds\footnote{The generating function for cumulants of $\delta(N_B-N_{\bar B})$ in a finite acceptance, $g_{net}$, reduces to the generating function for cumulants of $\delta(N_B+N_{\bar B})$ in the full system, Eq.~(\ref{eq:can_cumulant_gen_function}), in the limit $\alpha_B\to 1$ and $\alpha_{\bar B}\to -e^t$.}
\begin{eqnarray}\label{eq:generating-function}
    g_{net}(t)&=&\ln\left(\sum_{B_A} P_A(B_A)e^{B_A t}\right)\nonumber\\
    &=&\frac{B}{2}\left[\ln(q_1(t))-\ln(q_2(t))\right]+\ln\left\{I_B[2\, z\, \sqrt{q_1(t)\, q_2(t)}]\right\},
\end{eqnarray}
where $B$ is the total net-baryon number, $q_1(t)=1-\alpha_B+\alpha_B\, e^t$ and $q_2(t)=1-\alpha_{\bar B}+\alpha_{\bar B}\,e^{-t}$.

In order to compute the net-baryon number cumulants, we employ the formula of Fa\`{a} di Bruno (\ref{eq:faa-di-bruno-1})
to evaluate derivatives of $g_{net}(t)$.
For the first two terms in (\ref{eq:generating-function}), denote as $(a)$ and $(b)$, we choose $f(y)=\ln(y)$, $h(t)=q_1(t)$ and $f(y)=-\ln(y)$, $h(t)=q_2(t)$, as in~\cite{Braun-Munzinger:2020jbk}. Employing once more the Fa\`{a} di Bruno formula for functions of the form $\log(f(y))$~\cite{Comtet:1974}, we find a closed-form expression for the corresponding contribution to the net-baryon number cumulants~\cite{Braun-Munzinger:2020jbk}
\begin{eqnarray}\label{eq:cumulants-a-b}
    \kappa^{(a+b)}_n&=&\frac{B}{2}\Big(\kappa_{\rm Ber}^{(n)}(\alpha_B)+(-1)^{(n+1)}\kappa_{\rm Ber}^{(n)}(\alpha_{\bar B})\Big)\nonumber\\
    &\equiv&B\,k^{(n)}_+,
\end{eqnarray}
where $\kappa_{\rm Ber}^{(n)}(\alpha_B)$ is the $n$:th cumulant of the Bernoulli distribution (\ref{eq:Bernoulli-cumulant}), and the second line defines $k^{(n)}_+$.

In the evaluation of the derivatives of the last term in (\ref{eq:generating-function}), denoted by ${(c)}$, we deviate slightly from ~\cite{Braun-Munzinger:2020jbk}. We select $f(y)=\ln\big(I_B[2\,z\,e^y]\big)$ and $h(t)=\frac12\big(\ln(q_1(t))+\ln(q_2(t))\big)$ and evaluate the derivatives at $y=1$ and $t=0$, respectively. Using the fact that the derivatives of $f(y)$ are given by the cumulants $C_k$ (cf. (\ref{eq:can_cumulant_gen_function},\ref{eq:can-cumulant-Ntot})) and those of $h(t)$ by~\cite{Braun-Munzinger:2020jbk}
\begin{equation}
    h^{(n)}\equiv k_-^{(n)}=\frac12\Big(\kappa_{\rm Ber}^{(n)}(\alpha_B)-(-1)^{(n+1)}\kappa_{\rm Ber}^{(n)}(\alpha_{\bar B})\Big),
\end{equation}
we find that the corresponding contributions to the cumulants are given by
\begin{equation}\label{eq:kappa-c-cumulants}
    \kappa^{(c)}_n=\sum_{k=1}^n\,C_k\,B_{n,k}(k_-^{(1)},\dots ,k_-^{(n-k+1)}).
\end{equation}
Thus, the net-baryon cumulants in a subsystem are closely related to the cumulants of $\delta N^{tot}=\delta (N_B+N_{\bar B})$ in the full canonical system, where the net-baryon number is strictly conserved. Collecting all terms, we find a compact form for the net-baryon cumulants in a subsystem of a canonical system,
\begin{equation}\label{eq:cumulants-closed-form}
    \kappa_n=B\,k_+^{(n)}+\sum_{k=1}^n\,C_k\,B_{n,k}(k_-^{(1)},\dots ,k_-^{(n-k+1)}).
\end{equation}
We note that, using (\ref{eq:Nn-Ck-relation}), the cumulants $\kappa_n$ can be expressed in terms of the factorial cumulants $F_k$ rather than cumulants $C_k$. The resulting analytic expression for $\kappa_n$ is identical to the one obtained in~\cite{Braun-Munzinger:2020jbk} in terms of the functions $f^{(k)}(z)$, which, as previously mentioned, are equal to the factorial cumulants (\ref{eq:factorial-cumulants-tot}).

In appendix \ref{sect:App-c}, we obtain general relations of the net-baryon cumulants to the baryon and antibaryon cumulants in the subsystem. Specifically, we have,
\begin{equation}\label{eq:kappa-Cmn-text}
    \kappa_n=\sum_{i=0}^n\,\binom{n}{i}\,(-1)^{n-i}\,C^{(i,n-i)},
\end{equation}
where $\binom{n}{i}$ is a binomial coefficient. Furthermore, using (\ref{eq:cumulant-fac-cumulant-relation-subsystem-text}) one finds,
\begin{equation}\label{eq:cumulant-factorial-cumulant-relation}
    \kappa_n=\sum_{\substack{k_1,k_2\\ k_1+k_2\geq 0}}^{n}F^{(k_1,k_2)}\,\sum_{i=0}^n\binom{n}{i}\,(-1)^{n-i}\,S(i,k_1)\,S(n-i,k_2).
\end{equation}
For the first few cumulants, (\ref{eq:cumulant-factorial-cumulant-relation}) yields,
\begin{eqnarray}\label{eq:cumulants-from-can-cumulants}
    \kappa_1&=&F^{(1,0)}-F^{(0,1)},\nonumber\\
    \kappa_2&=&F^{(1,0)}+F^{(0,1)}+F^{(2,0)}+F^{(0,2)}-2\,F^{(1,1)},\nonumber\\
    \kappa_3&=&F^{(1,0)}-F^{(0,1)}+3\,(F^{(2,0)}-F^{(0,2)})\nonumber\\
    &-&3\,(F^{(2,1)}-F^{(1,2)})+F^{(3,0)}-F^{(0,3)},\nonumber\\
    \kappa_4&=&F^{(1,0)}+F^{(0,1)}+7\,(F^{(2,0)}+F^{(0,2)})+6\,(F^{(3,0)}+F^{(0,3)})\nonumber\\
    &-&6\,(F^{(2,1)}+F^{(1,2)})-4\,(F^{(3,1)}+F^{(1,3)})+6\,F^{(2,2)}-2\,F^{(1,1)}\nonumber\\
    &+&F^{(4,0)}+F^{(0,4)},\\
    \kappa_5&=&F^{(1,0)}-F^{(0,1)}+15\,(F^{(2,0)}-F^{(0,2)})+25\,(F^{(3,0)}-F^{(0,3)})\nonumber\\
    &+&10\,(F^{(4,0)}-F^{(0,4)})-15\,(F^{(2,1)}-F^{(1,2)})-20\,(F^{(3,1)}-F^{(1,3)})\nonumber\\
    &-&5\,(F^{(4,1)}-F^{(1,4)})+10\,(F^{(3,2)}-F^{(2,3)})+F^{(5,0)}-F^{(0,5)}\nonumber\\
    \kappa_6&=&F^{(1,0)}+F^{(0,1)}+31\,(F^{(2,0)}+F^{(0,2)})+90\,(F^{(3,0)}+F^{(0,3)})\nonumber\\
    &+&65\,(F^{(4,0)}+F^{(0,4)})+15\,(F^{(5,0)}+F^{(0,5)})-30\,(F^{(2,1)}+F^{(1,2)})\nonumber\\
    &-&80\,(F^{(3,1)}+F^{(1,3)})-45\,(F^{(4,1)}+F^{(1,4)})-6\,(F^{(5,1)}+F^{(1,5)})\nonumber\\
    &+&30\,(F^{(3,2)}+F^{(2,3)})+15\,(F^{(4,2)}+F^{(2,4)})-20\,F^{(3,3)}+30\,F^{(2,2)}\nonumber\\
    &-&2\,F^{(1,1)}+F^{(6,0)}+F^{(0,6)},\nonumber
\end{eqnarray}
When the factorial cumulants (\ref{eq:factorial-cumulants-closed-form}, \ref{eq:factorial-cumulants-3}) are substituted into (\ref{eq:cumulant-factorial-cumulant-relation}, \ref{eq:cumulants-from-can-cumulants}), the explicit expressions presented in~\cite{Braun-Munzinger:2020jbk} are recovered.

\section{Multi-particle correlations}
\label{Local_conservation_laws}
A precise interpretation of the fluctuation measurements requires considering not only the effects of global conservation laws but also local particle correlations that emerge from microscopic dynamics. 
These local correlations can be broadly categorized into two classes: correlations due to (i) repulsive and (ii) attractive interactions.

Repulsive interactions generally reduce the likelihood of particles appearing close to one another in phase space, which in turn leads to sub-Poissonian fluctuations, characterized by a reduced variance. Conversely, attractive interactions enhance the tendency of particles to aggregate, resulting in super-Poissonian fluctuations with an increased variance. Consequently, experimentally observed fluctuations are the result of a nontrivial interplay between global constraints and local correlations. Accurately quantifying and separating these effects is essential for isolating genuine dynamical signals, particularly those related to phase transitions and the conjectured QCD critical endpoint.

In the vicinity of a critical point, in particular the hypothetical QCD critical end point, correlations become long-ranged. This is manifested in a growing correlation length and accompanied by enhanced fluctuations. Experimental searches for the QCD critical point rely on detecting these enhanced fluctuations in a beam energy scan. Consequently, any viable theoretical model must account for both local correlations and global constraints in a consistent framework. In this study, we generate correlations in a phenomenological approach by introducing attractive and repulsive interactions. 

\subsection{Introducing Attractive and Repulsive Interactions}
The impact of local attractive correlations in rapidity space on fluctuation observables was examined in Ref.~\cite{Braun-Munzinger:2023gsd}. The events were generated employing a Metropolis algorithm with simulated annealing, using the cost function
\begin{equation}
\label{eq_cost_Metropolis}
\Delta_{n} = |\rho_{n} - \rho| - |\rho_{n-1} - \rho|.
\end{equation}
Here $\rho$ denotes the target (desired) value of the correlation coefficient, and $\rho_{n-1}$ and $\rho_n$ are the values of the correlation coefficient for two successive configurations of rapidity values in the iterative procedure of the Metropolis algorithm.

The correlation coefficient $\rho$ is defined as
\begin{equation}
\rho \equiv \frac{\text{Cov}(y_{1}, y_{2})}{\sigma_{y_{1}} \sigma_{y_{2}}} = \frac{\langle y_{1} y_{2} \rangle - \langle y_{1} \rangle \langle y_{2} \rangle}{\sigma_{y_{1}} \sigma_{y_{2}}},
\end{equation}
where the indices $y_{1}$ and $y_{2}$ refer to rapidity values of either a particle and its corresponding antiparticle, or of two particles. Here, $\text{Cov}(y_{1}, y_{2})$ denotes the covariance between the rapidity distributions of $y_{1}$ and $y_{2}$, while $\sigma_{y_{1}}$ and $\sigma_{y_{2}}$ represent the standard deviations of these distributions. A positive value for $\rho$ corresponds to attraction between the particles. 

The acceptance probability of a proposed change in rapidity space is given by
\begin{equation}
P_{n} = e^{-\Delta_{n}/\mathbb{T}},
\end{equation}
where $\mathbb{T}$ is an effective temperature parameter that is gradually reduced during the simulation, following the principles of the simulated annealing algorithm~\cite{Braun-Munzinger:2023gsd}. 

In this work, we generalize this method to encompass both repulsive and attractive interactions. To introduce correlations between random variables in a physically motivated and controllable way, we adopt a probabilistic framework, inspired by equilibrium statistical mechanics. Specifically, we define a joint probability density function for pairs (or more generally, sets) of random variables that obey the Boltzmann-Gibbs distribution:
\begin{equation}
P(y_{1},y_{1})=\frac{e^{-E(y_{1},y_{2})}}{Z},
%\label{netcumulants}
\end{equation}
where is the energy function encoding correlations between the random variables $y_{1}$ and $y_{2}$, while $Z$  is the partition function that ensures normalization of the probability distribution. With this form, generated configurations $\{y_{1},y_{2}\}$ corresponding to low-energy states are more probable, while those corresponding to high-energy configurations are less likely. This approach emulates the population of microstates in grand canonical systems, with the inverse temperature absorbed into the coefficients of the energy function. Here, the energy landscape assumes a pivotal role in shaping the emergent correlation structure.

To model the nature of correlations, we decompose the energy function \( E(y_1, y_2) \) into two components:

\begin{itemize}
    \item A \textbf{repulsive} term, which penalizes configurations where \( y_1 \) and \( y_2 \) are close to each other,
    \item An \textbf{attractive} term, which favors configurations where \( y_1 \) and \( y_2 \) are separated by a certain distance.
\end{itemize}

For the repulsive part, we use a Buckingham-type potential~\cite{Buckingham1938}:
\begin{equation}
\label{pot_r}
    E_r(y_1, y_2) = \alpha_r e^{-|y_1 - y_2| / \rho_r},
\end{equation}
where \( \alpha_r > 0 \) controls the strength of the repulsion, and \( \rho_r > 0 \) sets the interaction range. This short-range potential rapidly diminishes with increasing separation, effectively preventing particles from approaching each other too closely.

For the attractive part of the interaction, we employ a phenomenological confining potential of the power-law form:

\begin{equation}
\label{pot_a}
    E_a(y_1, y_2) = \alpha_a |y_1 - y_2|^{\beta_a},
\end{equation}
where \( \alpha_a > 0 \) denotes the strength of the attraction, and \( \beta_a > 0 \) controls the steepness or curvature of the potential. This potential ensures that the interaction energy increases progressively with separation \( |y_1 - y_2| \), effectively simulating a confining force.
The parameters \( \alpha_r, \rho_r, \alpha_a, \beta_a \) are determined  so that experimental results can be simulated using the Metropolis algorithm.

For clusters comprising more than two particles, we construct the total interaction energy as the sum over all pairwise interactions. For instance, in the case of a cluster composed of \( n \) particles with rapidity values \( y_1, y_2, \dots, y_n \), the total attractive interaction potential is given by the sum over all distinct pairs
\begin{equation}\label{eq:pot_a-n-part}
E_{a}^{\text{total}} = \sum_{1 \leq i < j \leq n} \alpha_a |y_i - y_j|^{\beta_a}.
\end{equation}

Similarly, multi-particle repulsion can be introduced by summing the repulsive potential over all pairs. Moreover, we extend the scheme presented in Ref.~\cite{Braun-Munzinger:2023gsd} to incorporate multi-particle correlations. This is achieved by introducing a modified version of the cost function $\Delta_{n}$, 
\begin{equation}
\label{pot_a_old}
    \Delta_{n}^{\mathrm{cluster}} =
    \sqrt{\sum_{1 \leq i < j \leq m}(\rho_{n}^{i,j} - \rho)^{2}}
    - \sqrt{\sum_{1 \leq i < j \leq m}(\rho_{n-1}^{i,j} - \rho)^{2}},
\end{equation}
where $\rho$ again denotes the target value of the correlation coefficient, $\rho_{n}^{i,j}$ and $\rho_{n-1}^{i,j}$ are the values of the correlation coefficients for two successive configurations of rapidity values for pairs $i$ and $j$ within the cluster, and $m$ is the number of particles in the cluster.  
In the following, we restrict ourselves to clusters composed of either two or three particles.  

For attractive interactions, we employ two alternative approaches:  
(i) introducing a phenomenological potential as defined in Eq.~\ref{pot_a}, and  
(ii) using the cost function defined in Eq.~\ref{pot_a_old}.  
The results from both approaches are compared with experimental data from HADES and STAR, as discussed below.

\section{Comparison with experimental results}
\label{Comparison_with_experimental_results}

In the subsequent subsections, we explore the rapidity and energy dependence of the ratios of proton factorial cumulants\footnote{In the literature, the factorial cumulant of order $n$ for a multiplicity distribution of particle type $q$ is sometimes denoted as $C_{n}(q)$~\cite{Holzmann:2024wyd,Nabroth:2025qm}. Moreover, the STAR experiment employs the notation $\kappa_{n}(q)$ for factorial cumulants, while for ordinary cumulants, it uses $C_{n}(q)$~\cite{STAR:2025zdq}.}, $\Fpr{n}$. Specifically, we confront results for factorial cumulants up to fourth order, obtained within the canonical ensemble encompassing both attractive and repulsive proton--proton interactions, with the data obtained by the STAR and HADES experiments. 

The relation of the factorial cumulants of the proton number distribution $\Fpr{n}$  to the corresponding ordinary proton cumulants $C^{(n,0)}$  follows from Eq.~\ref{eq:fac-cumulant-cumulant-relation-subsystem}
\begin{eqnarray}\label{eq:fac-cumulant-cumulant-no-antibaryons}
    \Fpr{1}&=&C^{(1,0)}, \nonumber\\
    \Fpr{2}&=&-C^{(1,0)}+C^{(2,0)},\\
    \Fpr{3}&=&2\,C^{(1,0)}-3\,C^{(2,0)}+C^{(3,0)}
    ,\nonumber\\
    \Fpr{4}&=&-6\,C^{(1,0)} +11\,C^{(2,0)}-6\,C^{(3,0)}+C^{(4,0)}\nonumber.
\end{eqnarray}

The input parameters for our calculations, including the acceptance criteria for protons and antiprotons, and the average baryon and antibaryon yields in full phase space, were determined in Ref.~\cite{Braun-Munzinger:2020jbk}. Given that the STAR measurements are restricted to midrapidity, they do not provide baryon rapidity distributions in full phase space, essential in our approach. Consequently, we perform the analysis at $\sqrt{s_{NN}} = 8.8$, 17.3, and 62.4~GeV, using as input net-proton and net-baryon rapidity distributions obtained by the NA49 and BRAHMS collaborations (see Ref.~\cite{Braun-Munzinger:2020jbk} for details).

\begin{figure}[htb]
\centering
 \includegraphics[width=1.\linewidth,clip=true]{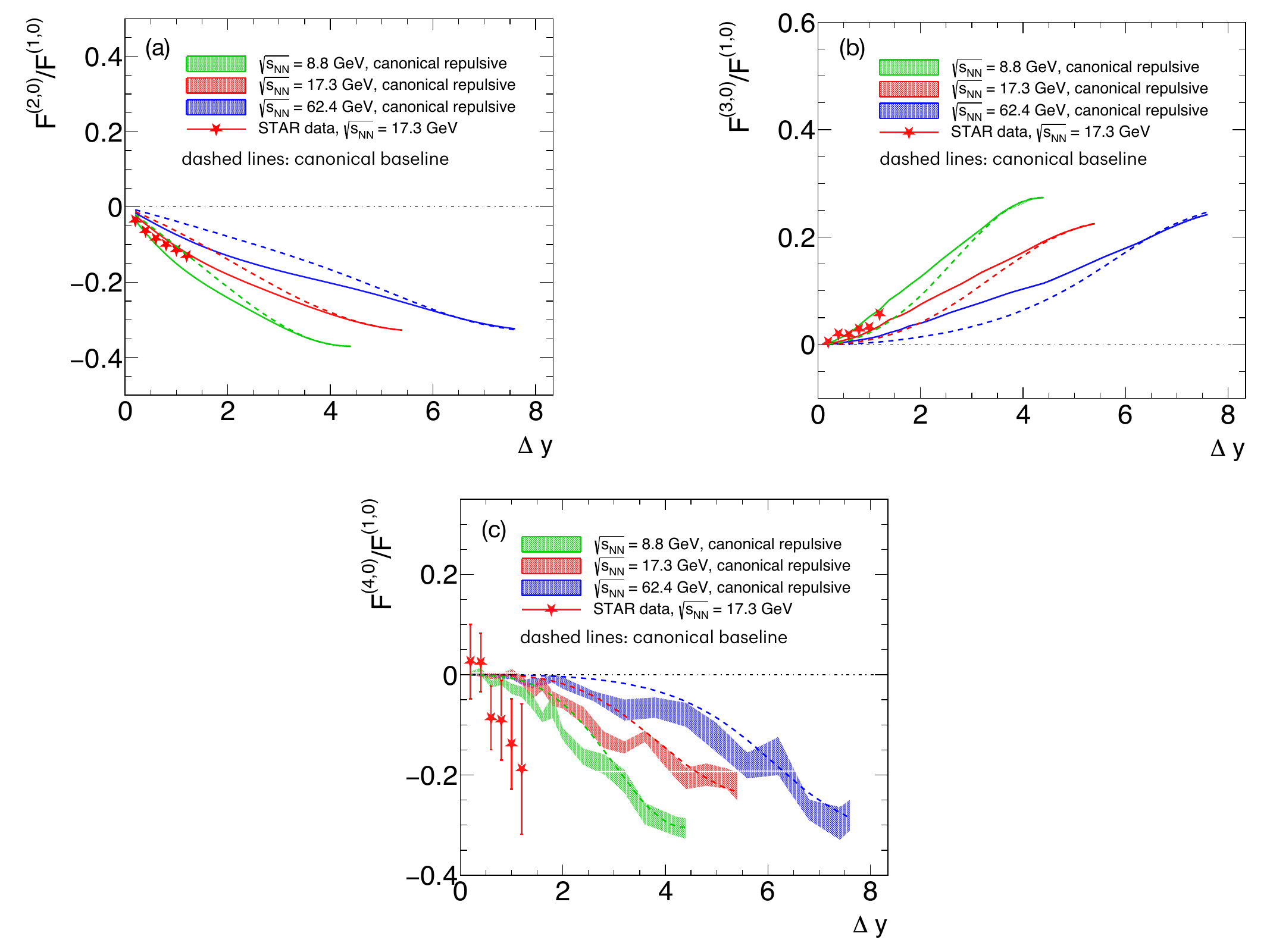}
 \caption{Panels (a), (b), and (c) show the dependence on the  rapidity gap of the factorial cumulant ratios $\Fnorm{2}$, $\Fnorm{3}$, and $\Fnorm{4}$ of proton number distributions for collision energies $\snn = 8.8$, 17.3, and 62.4~GeV. Colored solid and hashed lines represent results from Monte Carlo simulations with 150 million events that include repulsive interactions between proton pairs. The width of the hashed lines indicates statistical uncertainties. Dashed colored lines correspond to analytical calculations that account for global baryon number conservation only (canonical baseline). Also shown are the data of the STAR experiment at $\snn = 17.3$~GeV (red symbols).} 
\label{fig_fact_cumulants_Energy}
\end{figure} 

At lower energies, $\sqrt{s_{NN}} = 2.55$~GeV, we use measured proton rapidity distributions in Ag--Ag collisions from the HADES experiment~\cite{Nabroth:2025qm}. For these collisions, we also utilize the mean number of protons as a function of the rapidity interval, as well as the average number of wounded nucleons, determined by the HADES collaboration within the wounded nucleon model~\cite{Nabroth:2025qm, HADES:2017def}. Furthermore, we present recent results for Au--Au collisions obtained by the STAR collaboration in the Fixed-Target Mode (FXT STAR) at $\sqrt{s_{NN}} = 3.0$, 3.2, 3.5, and 3.9~GeV per nucleon pair~\cite{Zachary:2025qm}.

\subsection{Rapidity Dependence of the Factorial Cumulant Ratios}
\label{Predictions}

Fig.~\ref{fig_fact_cumulants_Energy} displays $\Fnorm{2}$, $\Fnorm{3}$, and $\Fnorm{4}$ as functions of the rapidity coverage for $\sqrt{s_{\mathrm{NN}}} = 8.8$, 17.3, and 62.4~GeV. The colored solid and hashed lines show the result of canonical simulations with 150 million events, accounting for repulsive proton--proton interactions. The dashed lines show the corresponding ratios obtained using  (\ref{eq:factorial-cumulants-closed-form}), i.e., taking only global conservation laws into account. In the following, we refer to the latter as the \emph{canonical baseline}. The width of the hashed lines indicates statistical uncertainties estimated using the sub-sampling method~\cite{Anticic:2013htn, ALICE:2017jsh}. 
\begin{figure}[htb]
\centering
 \includegraphics[width=1.\linewidth,clip=true]{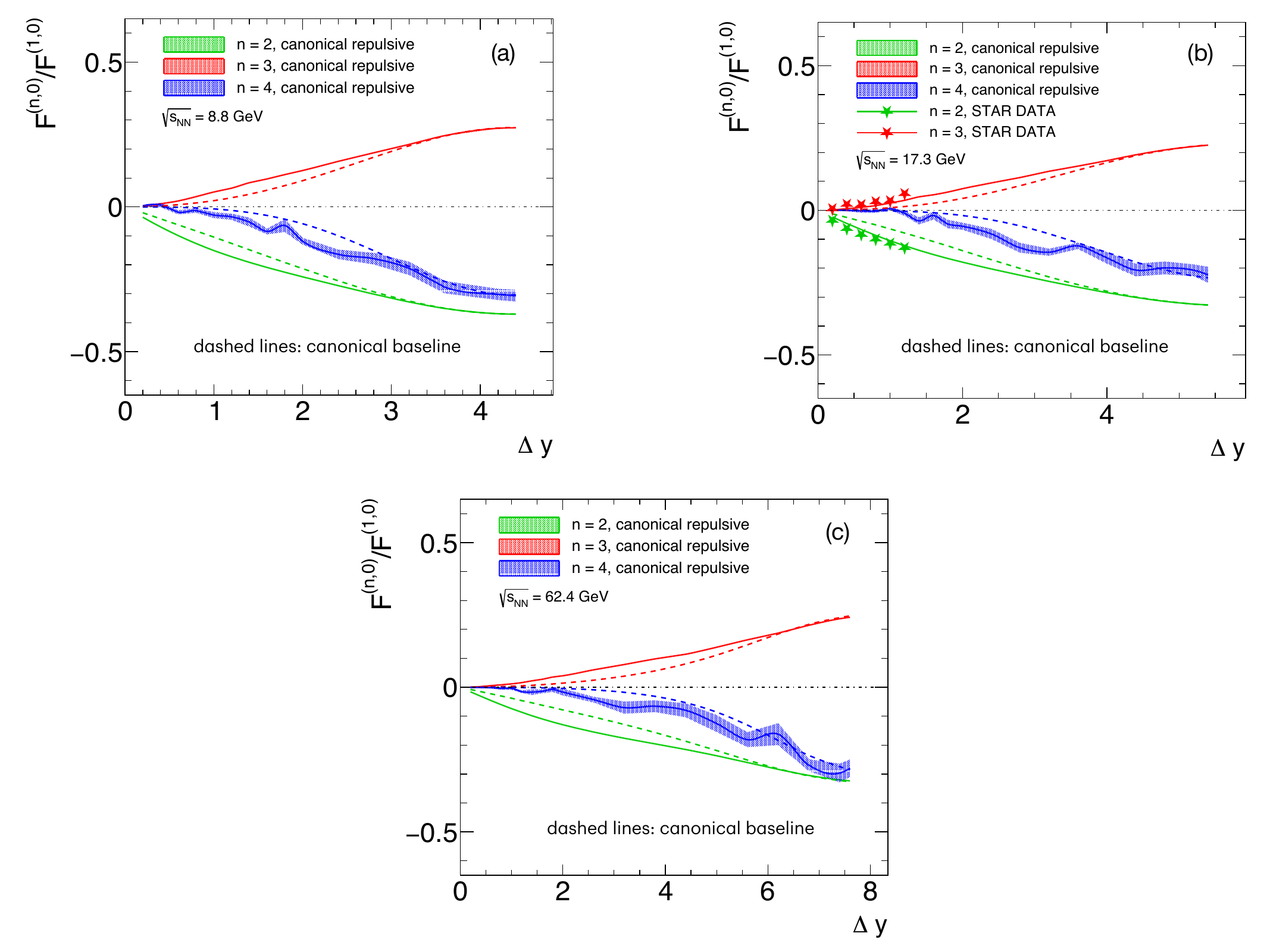}
 \caption{The dependence on the  rapidity gap of factorial cumulant ratios for proton number distributions at $\snn = 8.8$, 17.3, and 62.4 GeV is presented in panels (a), (b), and (c), respectively. Colored solid and hashed lines represent results from Monte Carlo simulations with 150 million events that include repulsive interactions between proton pairs. The width of the hashed lines indicates statistical uncertainties. Dashed colored lines correspond to analytical calculations that account for global baryon number conservation only (canonical baseline). The data obtained by STAR collaboration at $\snn = 17.3$~GeV are shown in panel (b) as red (n=3) and green (n = 2) symbols.} 
\label{fig_fact_cumulants_All}
\end{figure} 
\begin{figure}[htb]
\centering
 \includegraphics[width=1.\linewidth,clip=true]{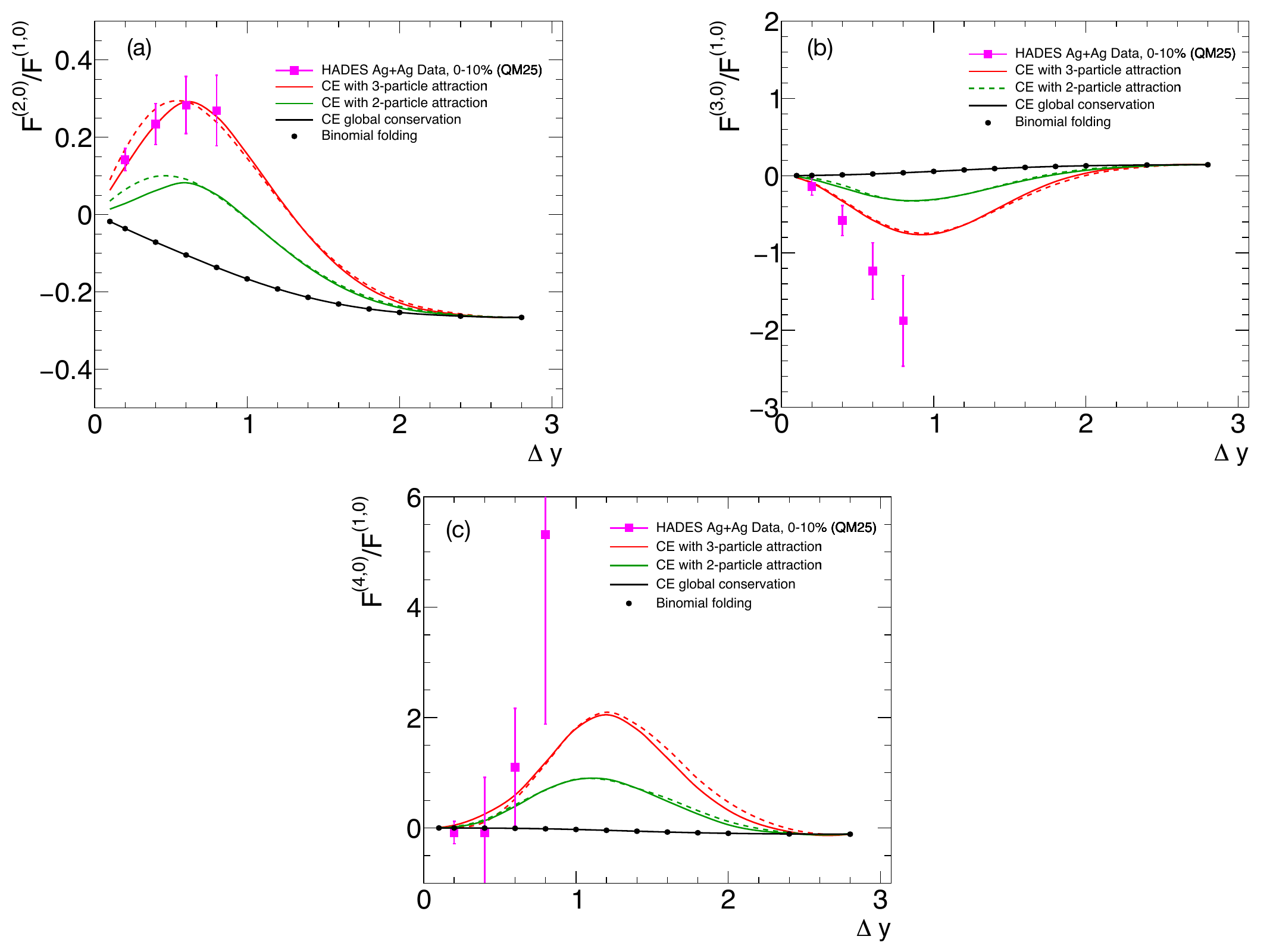}
 \caption{The dependence on the rapidity gap of the factorial cumulant ratios 
$\Fnorm{2}$, $\Fnorm{3}$, and $\Fnorm{4}$ 
for proton number distributions in central Ag--Ag collisions, 
as measured by the HADES experiment (purple solid squares). 
The colored solid lines represent results from Monte Carlo simulations with 100 million events within the canonical ensemble framework, including attractive correlations among three protons (red) and two protons (green), computed using Eq.~\ref{eq:pot_a-n-part}. The corresponding simulations based on Eq.~\ref{pot_a_old} are shown as dashed lines. The black line denotes analytical calculations including only global baryon
number conservation (canonical baseline), i.e., without local interactions. Black solid circles represent the results of binomial folding of the baryon number factorial cumulants in full phase space (cf. Eqs.~\ref{eq:asyptbaryons}~\ref{eq:asyptprotos}).} 
\label{fig_fact_cumulants_HADES}
\end{figure} 
For small acceptances, both results (canonical ensemble with repulsive interactions and the canonical baseline) approach zero, the small-number Poisson limit, as expected. At full acceptance, the two schemes yield the same (non-zero) results. Clearly, in this limit it is irrelevant whether two protons are correlated or not, since they cannot leave the acceptance window. The limiting values for the cumulant ratios are then determined only by global baryon number conservation. By contrast, at intermediate acceptances, two-proton correlations clearly affect the results, leading to clear differences for the three ratios considered. 

Furthermore, in Fig.~\ref{fig_fact_cumulants_Energy} the STAR BES-II results at $\snn = 17.3$~GeV are shown as red symbols~\cite{Huang:2025qm}. Clearly, the STAR data on $\Fnorm{2}$ lie systematically below the canonical baseline (red dashed line), whereas for $\Fnorm{3}$ the situation is reversed. The inclusion of repulsive interactions between protons significantly improves the agreement with the data for both ratios, highlighting the importance of incorporating such interactions in the modeling. This agreement is very good for $\Fnorm{2}$ and $\Fnorm{3}$, while for $\Fnorm{4}$ the model results are, given the large uncertainties, consistent with the data\footnote{Note that all ratios are computed with the same set of model parameters.}. Moreover, introducing attractive interactions rather than repulsive ones, would shift these ratios in the opposite direction relative to the canonical baseline (see below for details). This is clearly disfavored by the measurements.

In Fig.~\ref{fig_fact_cumulants_All}, we present the cumulant ratios from a complementary perspective: the ratios $\Fnorm{n}$ with $n \in [2,4]$ are shown for three representative collision energies, $\snn = 8.8$, 17.3, and 64 GeV, in panels (a), (b), and (c), respectively. Across all energies, both $\Fnorm{2}$ and $\Fnorm{4}$ take negative values, while $\Fnorm{3}$ remains positive. We further observe that higher-order cumulant ratios, such as $\Fnorm{4}$, require larger rapidity intervals to display deviations from the Grand Canonical Ensemble (GCE) baseline, indicated by the dash-dotted black line at zero. In panel~(b), the corresponding STAR BES-II measurements~\cite{Huang:2025qm} are shown as solid star symbols. As discussed above, they are consistent with the model results that account for repulsive interactions (solid red and green lines).

The solid purple squares in Fig.~\ref{fig_fact_cumulants_HADES} represent the rapidity dependence of the proton factorial cumulant ratios, obtained by the HADES Collaboration for Ag--Ag collisions at $\sqrt{s_{NN}} = 2.55$~GeV~\cite{Nabroth:2025qm}, using a recently proposed method for the analysis of fluctuation data~\cite{Rustamov:2024hvq}. 
The solid lines correspond to results of Monte Carlo simulations that incorporate three-proton (red solid lines) and two-proton (green solid lines) attractive interactions~\footnote{The description of the data on  $\Fnorm{2}$ with two-proton attraction could be improved by tuning the interaction strength. Thereby one could obtain reasonable agreement with the data at $\Delta y=0.6$ and $0.8$. However, the model would then overshoot the two data points at a smaller rapidity gap by more than two standard deviations. Thus, we find that the three-proton attractive correlation yields a much better description of the shape of the data.}, computed using Eq.~\ref{pot_a}. The corresponding results for attraction between two and three protons using Eq.~\ref{pot_a_old} are presented by the dashed lines. In general, the results of the two schemes are very similar. 

The black solid lines show the canonical baseline, i.e., without local interactions. The black solid circles are obtained by taking the baryon number in full phase space, calculating its cumulants, and folding them with the acceptance using binomial efficiencies~\cite{Nonaka:2017kko}. The acceptance is computed as the ratio of protons inside the acceptance to the total number of baryons in full phase space. As expected, baryon number factorial cumulants folded with binomial efficiencies are equivalent to the canonical baseline results. This should be the case at low collision energies, where the number of anti-baryons is negligible and hence the total number of baryons is fixed and hence does not fluctuate (see Appendix C of Ref.~\cite{Braun-Munzinger:2020jbk}). Using this condition, one obtains the asymptotic values for the factorial cumulants of baryons in the full phase space (cf. Eq.~\ref{eq:fac-cumulant-cumulant-no-antibaryons})
\begin{equation}
 \label{eq:asyptbaryons}
    \frac{\Fpr{2}(B^{4\pi})}{\Fpr{1}(B^{4\pi})} \rightarrow -1, \quad
    \frac{\Fpr{3}(B^{4\pi})}{\Fpr{1}(B^{4\pi})} \rightarrow 2, \quad
    \frac{\Fpr{4}(B^{4\pi})}{\Fpr{1}(B^{4\pi})} \rightarrow -6.
\end{equation}

By folding these cumulant values with the binomial efficiency, defined as the ratio of the number of protons to the total number of baryons in the full phase space, we obtain the asymptotic values of the proton cumulant ratios, corresponding to full rapidity coverage and the final experimental transverse momentum acceptance
\begin{equation}
\label{eq:asyptprotos}
    \frac{\Fpr{2}}{\Fpr{1}} \rightarrow -0.27, \quad
    \frac{\Fpr{3}}{\Fpr{1}} \rightarrow 0.14, \quad
    \frac{\Fpr{4}}{\Fpr{1}} \rightarrow -0.11.
\end{equation}
As shown in Fig.~\ref{fig_fact_cumulants_HADES}, the asymptotic values of the cumulant ratios of the proton number are reached in the full rapidity space, both with and without local interactions. As noted above, local correlations play no role in this limit. Moreover, at low energies, where the number of antibaryons is negligible, the ratios in full phase space are determined only by the proton-to-baryon ratio. Conversely, in the limit of vanishing acceptance, all factorial cumulant ratios approach the Poissonian limit, i.e., zero.

The results presented in Figs. \ref{fig_fact_cumulants_Energy}, \ref{fig_fact_cumulants_All} and \ref{fig_fact_cumulants_HADES} clearly indicate that the nature of proton–proton interactions evolves from being predominantly repulsive at higher energies to predominantly attractive at lower energies. 

\begin{figure}[htb]
\centering
\includegraphics[width=1.\linewidth,clip=true]{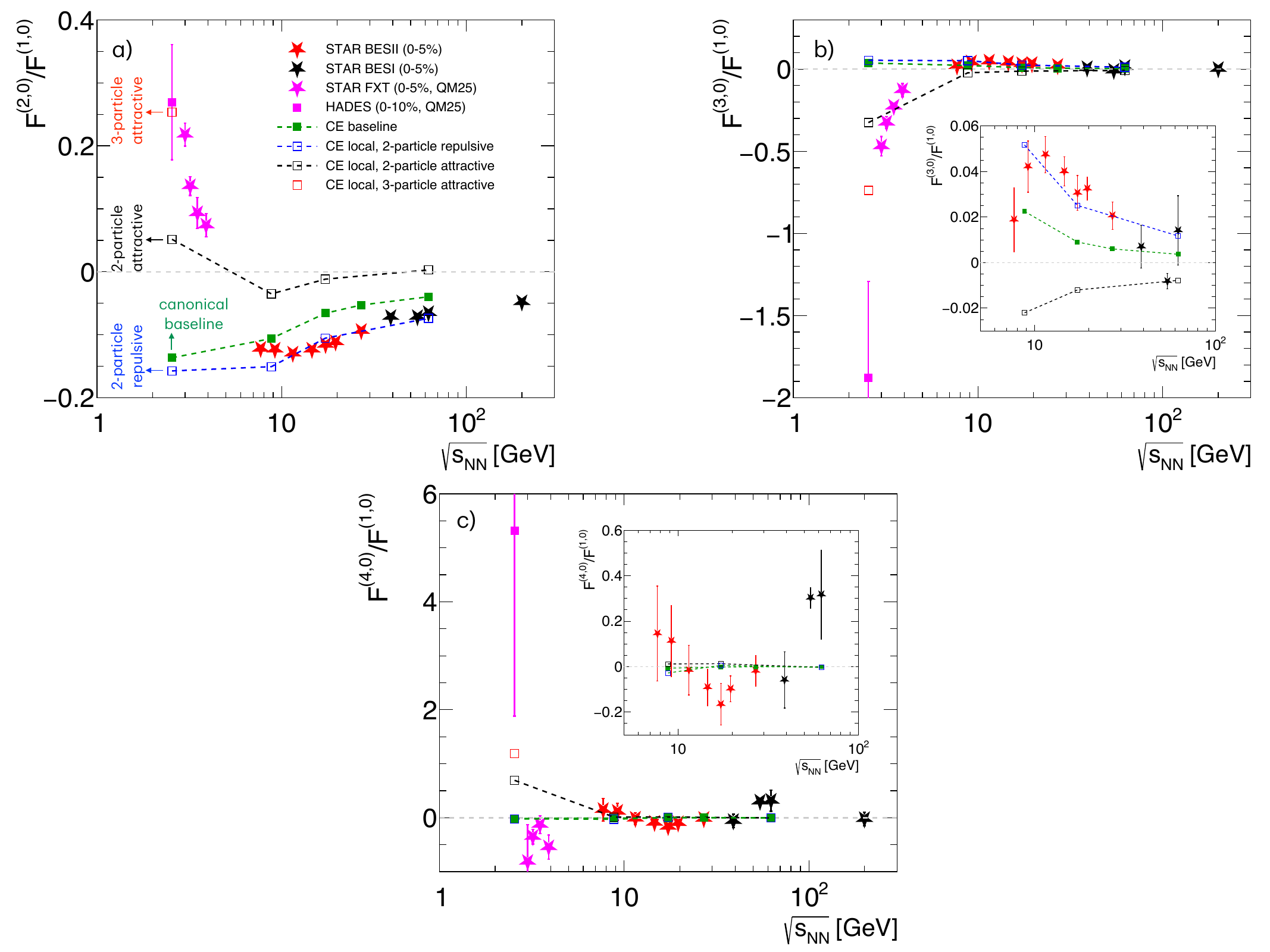} 
 \caption{Ratios of proton factorial cumulants, $\Fnorm{2}$, $\Fnorm{3}$, and $\Fnorm{4}$, are shown in panels (a), (b), and (c). The solid black and red stars denote experimental results for central Au–Au collisions from STAR BES-I and BES-II, respectively. The solid purple boxes show the HADES measurement for Ag–Ag collisions, while the purple stars correspond to STAR FXT measurements. Solid green boxes indicate our analytical calculations incorporating global baryon number conservation (canonical baseline). Open blue boxes represent simulation results including local two-proton repulsive correlations, whereas open black boxes account for local two-proton attractive correlations. The open red boxes at the HADES energy show simulation results that include three-proton attractive interactions.} 
\label{fig_fact_cumulants}
\end{figure} 

\subsection{Energy Dependence of the Factorial Cumulant Ratios}

In Fig.~\ref{fig_fact_cumulants} we compare the energy dependence (excitation function) of the factorial cumulant ratios of proton number distributions with our model results. Black stars indicate STAR results from central Au--Au collisions in the Beam Energy Scan I (BES-I) program~\cite{STAR:2021iop}, at \snn = 54.4, 62.4 and 200 GeV, while red stars correspond to  results from BES-II at \snn = 7.7, 9.2, 11.5, 14.6, 17.3, 19.6 and 27 GeV~\cite{STAR:2021iop, Huang:2025qm}. Both BES-I and BES-II data represent the $5\%$ most central Au--Au collisions, analyzed in the transverse momentum interval $0.4 < p_{T}~[\mathrm{GeV}/c] < 2$ and rapidity interval $|y_{\mathrm{cm}}| < 0.5$. Solid purple stars represent STAR FXT data~\cite{STAR:2021iop,Zachary:2025qm} for the $5\%$ most central Au--Au collisions in the $-0.5 < y_{\mathrm{cm}} < 0$ rapidity interval, with the same transverse momentum coverage as in the BES-I/BES-II data. The solid purple boxes denote preliminary HADES measurements for the $10\%$ most central Ag--Ag collisions at \snn = 2.55 GeV, within the acceptance range $0.4 < p_{T}~[\mathrm{GeV}/c] < 1.6$ and $|y_{\mathrm{cm}}| < 0.4$~\cite{Nabroth:2025qm}. 

The corresponding calculations of different baselines include the experimental acceptance criteria in both transverse momentum and rapidity.

The analytical baseline calculations, which account exclusively for global baryon number conservation (canonical baseline), are shown as solid green boxes.
For $\Fnorm{2}$, the STAR measurements at intermediate collision energies lie systematically below the canonical baseline, whereas the HADES result is significantly above it. In contrast, for $\Fnorm{3}$, the STAR results remain consistently above the canonical baseline (see insert in panel (b)), while the HADES measurements are consistently below. This opposing behavior when comparing low- and high-energy collisions suggests that the deviations from the canonical baseline in the two energy regimes originate from different underlying physical mechanisms. A similar pattern is observed for $\Fnorm{4}$, albeit with larger uncertainties.

At the highest energies, the STAR measurements tend to converge toward the canonical baseline, a trend that can be largely attributed to limited experimental acceptance.\footnote{The STAR measurements are performed within a fixed rapidity window of $|y| < 0.5$ for (anti-)protons, independent of collision energy. As the rapidity distribution broadens with increasing energy, the acceptance fraction decreases, leading to an apparent suppression of fluctuations.}

Overall, these findings indicate that the conventional canonical ensemble framework, which includes only baryon number conservation, is insufficient to quantitatively reproduce the STAR and HADES results, motivating the incorporation of additional interaction effects in the theoretical modeling.

\begin{figure}[htb]
\centering
 \includegraphics[width=.5\linewidth,clip=true]{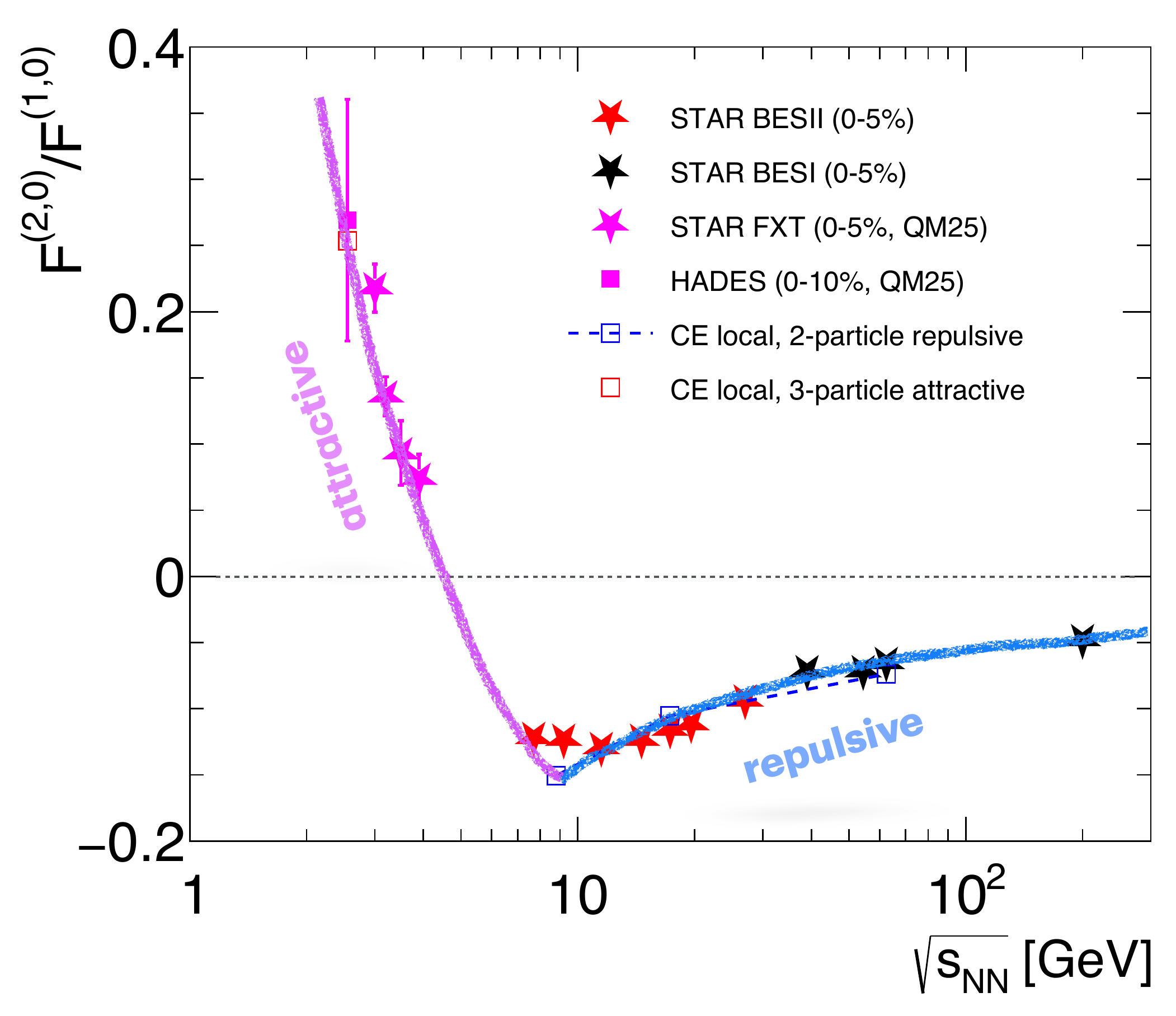}
 \caption{The transition from attractive to repulsive correlations with increasing collision energy for the ratio $\Fnorm{2}$. The solid purple line, connecting the calculation at HADES energy that includes three-particle attractive interactions (open red box) with calculations that account for two-particle repulsive interactions (open blue boxes)  is drawn to guide the eye. The filled symbols represent experimental measurements, also shown in panel (a) of Fig.~\ref{fig_fact_cumulants_HADES}.} 
\label{fig_attractive-repulsive}
\end{figure} 

The corresponding results, accounting for repulsive interactions between pairs of protons (two-particle repulsion), are indicated by open blue boxes in Fig.~\ref{fig_fact_cumulants}. These calculations utilize the energy function $E_{r}(y_{1},y_{2})$ defined in Eq.~\ref{pot_r}. The inclusion of repulsive interactions significantly improves the agreement with the STAR BES-II data. Nonetheless, deviations persist at lower energies: while low-energy BES-II data still show moderate disagreement with the baseline including repulsion, the discrepancy is more pronounced in the HADES and STAR FXT data, the latter indicated by purple star symbols.  

We also note that the $\Fnorm{2}$ and $\Fnorm{3}$ results from STAR FXT are consistent with those from HADES; however, this agreement should be interpreted with caution, as the analyses are performed in different acceptances. Moreover, while the STAR FXT data correspond to Au--Au collisions with an average number of wounded nucleons of $N_{W} \approx 320$ for the selected centrality, the HADES data are for Ag--Ag collisions with $N_{W} \approx 165$. Therefore, to draw firm conclusions, baselines for the STAR FXT data must also be calculated. At present, our low-energy baseline calculations are available only for the HADES data.

The observed behavior strongly suggests the need to incorporate attractive interactions among protons at low energies, since repulsive interactions alone drive the ratios in the opposite direction further away from the data.  

Simulations including pairwise attractive correlations are represented by open black boxes in Fig.~\ref{fig_fact_cumulants}. At the HADES energy, the result with three-proton (rather than two-proton) attraction are shown by the open red boxes. The latter provides a better description of the HADES data, as discussed above (see Fig.~\ref{fig_fact_cumulants_HADES} and related discussion).  

The parameters for two-particle repulsive interactions (Eq.~\ref{pot_r}) were determined with the Metropolis algorithm under the condition that the STAR BESII 
results are satisfactorily reproduced. This procedure yields $\alpha_{r} = 35$ and $\rho_{r} = 1$, which are used independently of the collision energy. At the HADES 
energy, the parameters for two- and three-particle attractive interactions 
(Eq.~\ref{eq:pot_a-n-part}) were obtained as $\alpha_{a} = 1000$ and $\beta_{a} = 1$. We also tested the earlier parametrization (Eq.~\ref{pot_a_old}) for the HADES data, which describes the measurements with a correlation coefficient of $\rho = 0.8$. Applying 
Eq.~\ref{eq:pot_a-n-part} with the above parameters yields a nearly identical value, $\rho \approx 0.8$, indicating that both formulations provide consistent results, 
as shown also in Fig.~\ref{fig_fact_cumulants_HADES}.

The energy dependence of the factorial cumulant ratios reveals a compelling indication for an interplay between repulsive and attractive interactions. Our calculations show that at high collision energies, repulsive correlations dominate, whereas at lower energies, attractive correlations become increasingly important. This behavior is schematically illustrated in Fig.~\ref{fig_attractive-repulsive}. Such a trend is reminiscent of the dynamics in classical systems described by the Van der Waals equation of state, where the competition between short-range repulsion and long-range attraction governs the liquid--gas phase transition.

\section{Summary and conclusions} 
\label{Summary}
The main objective of these studies was to calculate fluctuations of the multiplicity distributions of baryons, antibaryons, as well as their sum and difference in a canonical ensemble. 
We have focused on cumulants and factorial cumulants, assuming that   baryons are  correlated  by exact conservation of the net-baryon number in the  full system. 
To this end we have adopted a statistical model of baryons and antibaryons in the spirit of the S-matrix approach,  where  to leading order in the fugacity expansion, the  partition function takes the form of an ideal gas.    

We have derived analytic expressions 
for the cumulants and factorial cumulants of the baryon and antibaryon numbers in the canonical ensemble. Moreover, based on the generalized Fa\`{a} di Bruno formula for computing the derivatives of composite functions, we obtained closed-form results for cumulants and factorial cumulants of arbitrary order in a subsystem constrained
by exact conservation of the net-baryon number in the full system. 
We also presented general relations, independent of assumptions on the partition function, between cumulants and factorial cumulants as well as between net-baryon-number fluctuations in a subsystem and those of the baryon and antibaryon numbers of the full system. 

One of the essential novelties of this contribution is the inclusion of correlations, modelled by introducing attractive and repulsive interactions among multiparticle systems, while preserving baryon number conservation in the full phase space.

By comparing these developments to recent measurements from the \textsc{STAR} BES-II and \textsc{HADES} experiments, we arrive at an important conclusion: to describe the energy dependence of proton factorial cumulants over a broad range of collision energies, from $\sqrt{s_{NN}} = 17.3$\,GeV down to $\sqrt{s_{NN}} = 2.5$\,GeV, it is essential to account for correlations among protons. The consideration of global baryon number conservation alone is insufficient to reproduce the experimental observations. 

The inclusion of correlations, also successfully describes more differential measurements, such as the rapidity coverage dependence of proton factorial cumulants reported by \textsc{STAR} and \textsc{HADES}.

Furthermore, we establish that at higher energies, the dominant contribution arises from repulsive two-particle interactions between protons. In contrast, at lower energies, our results indicate the necessity of attractive three-particle interactions. This observation suggests that the nature of baryon interactions evolves with collision energy, from predominantly repulsive at high energies to attractive at low energies.

This behavior may represent an indication of a first-order phase transition in strongly interacting QCD matter, in analogy with the dynamics of the van der Waals equation of state, which modifies the ideal gas law by incorporating short-range repulsion and long-range attraction. The fact that the temperatures probed in these collisions are above 60 MeV, indicates that these fluctuation patterns are possibly connected  with the conjectured chiral transition of QCD rather than the gas-liquid transition of nuclear matter.

\section{Acknowledgments}
We are grateful for stimulating discussions with Mesut Arslandok, Peter Braun-Munzinger, Johanna Stachel and Joachim Stroth. Valuable input from Nu Xu, Yige Huang, Marvin Nabroth regarding recent STAR and HADES measurements is also appreciated. K.R. acknowledges the support by the Polish National Science Centre (NCN) under OPUS Grant No. 2022/45/B/ST2/01527 and by the Polish Ministry of Science and Higher Education.

\appendix
\section{Multivariate Bell polynomials}\label{sec:app-a}

In order to express the derivatives of a composite function of several variables, like $g_{fc}(x,\bar{x})$ in (\ref{eq:generating-function-fact-cum}), in a form analogous to the Fa\`{a} di Bruno formula one needs a generalization of the Bell polynomials~\cite{Bell:1927}.

The multivariate partial Bell polynomials can be obtained from a generalization of the generating function for the standard partial Bell polynomials~\cite{Schumann:2019xy}
\begin{equation}\label{eq:gen-function-Bell}
    \Phi(t,u)=\exp\left(u\sum_j x^{(j)}\frac{t^j}{j!}\right)=\sum_{n\geq k\geq 0}B_{n,k}(x^{(1)},x^{(2)},\dots,x^{(n-k+1)})\,\frac{t^n}{n!}\,u^k.
\end{equation}
The regular partial Bell polynomials are given by
\begin{equation}\label{eq:partial-Bell}
    B_{n,k}(x^{(1)},x^{(2)},\dots,x^{(n-k+1)})=\frac{1}{k!}\frac{\partial^n}{\partial t^n}\,\frac{\partial^k}{\partial u^k}\,\Phi(t,u)|_{t=u=0}.
\end{equation}
The Bell polynomials enter the Fa\`{a} di Bruno formula (\ref{eq:faa-di-bruno-1}) for the derivatives of a composite function $f(g(z))$,
Below we present the generalizations of (\ref{eq:faa-di-bruno-1}) to composite functions of the form $f(g(z_1,z_2))$ and $f(g_1(z),g_2(z))$.

In Ref.~\cite{Schumann:2019xy} the generating function for the general case of arbitrary dimensions of $z$ and $g$ is given. In order to keep the notation transparent, we give the generating function for two-dimensional $\vec{z}=(z_1,z_2)$ and $\vec{g}=(g_1,g_2)$,
\begin{eqnarray}\label{eq:gen-function-multiv-Bell}
    \Phi(t_1,t_2,u_1,u_2)&=&\exp\left(\sum_{\substack{j_1,j_2\\ j_1+j_2>0}}^\infty \big[u_1*x_1^{(j_1,j_2)}+u_2*x_2^{(j_1,j_2)}\big]\frac{t_1^{j_1}\,t_2^{j_2}}{j_1!\,j_2!}\right)\\
    &=&\sum_{\substack{n_1,n_2,k_1,k_2\\ n_1+n_2\geq k_1+k_2\geq 0}}B_{\{n_1,n_2\},\{k_1,k_2\}}\big(\{\vec{x}^{(i,j)}\}\big)\,\frac{t_1^{n_1}\,t_2^{n_2}}{n_1!\,n_2!}\,(u_1^{k_1}+u_2^{k_2}).\nonumber
\end{eqnarray}
The generalized Bell polynomials are then given by
\begin{eqnarray}\label{eq:gen-Bell-22}
    B_{n_1,n_2;k_1,k_2}(\{\vec{x}^{(i,j)}\})&=&\frac{1}{k_1!\,k_2!}\\
    &\times&\frac{\partial^{n_1}}{\partial t_1^{n_1}}\frac{\partial^{n_2}}{\partial t_2^{n_2}}\frac{\partial^{k_1}}{\partial u_1^{k_1}}\frac{\partial^{k_2}}{\partial u_2^{k_2}}\,\Phi(t_1,t_2,u_1,u_2)|_{t_1=t_2=u_1=u_2=0}.\nonumber
\end{eqnarray}

Now, for computing the factorial cumulants, we need the case with vector argument $\vec{z}=(z_1,z_2)$ and a scalar function $g(z_1,z_2)$,
\begin{eqnarray}\label{eq:gen-function-multiv-Bell-21}
    \Phi(t_1,t_2,u)&=&\exp\left(u\,\sum_{\substack{j_1,j_2\\ j_1+j_2>0}}^\infty x^{(j_1,j_2)}\frac{t_1^{j_1}\,t_2^{j_2}}{j_1!\,j_2!}\right)\\
    &=&\sum_{\substack{n_1,n_2,k\\ n_1+n_2\geq k\geq 0}}B_{n_1,n_2;k}\big(\{x^{(i,j)}\}\big)\,\frac{t_1^{n_1}\,t_2^{n_2}}{n_1!\,n_2!}\,u^{k}.\nonumber
\end{eqnarray}
The corresponding partial Bell polynomials are given by,
\begin{eqnarray}\label{eq:gen-Bell-21}
    B_{n_1,n_2;k}(\{x^{(i,j)}\})&=&\frac{1}{k!}\,\frac{\partial^{n_1}}{\partial t_1^{n_1}}\frac{\partial^{n_2}}{\partial t_2^{n_2}}\frac{\partial^{k}}{\partial u^{k}}\,\Phi(t_1,t_2,u)|_{t_1=t_2=u=0}.
\end{eqnarray}
The first few generalized Bell polynomials are
\begin{eqnarray}\label{eq:app-7}
    B_{1,0;1}(\{x^{(i,j)}\})&=&x^{(1,0)}\,,\nonumber\\
    B_{0,1;1}(\{x^{(i,j)}\})&=&x^{(0,1)}\,,\nonumber\\
    B_{2,0;1}(\{x^{(i,j)}\})&=&x^{(2,0)}\,,\nonumber\\
    B_{2,0;2}(\{x^{(i,j)}\})&=&(x^{(1,0)})^2\,,\\
    B_{0,2;1}(\{x^{(i,j)}\})&=&x^{(0,2)}\,,\nonumber\\
    B_{0,2;2}(\{x^{(i,j)}\})&=&(x^{(0,1)})^2\,,\nonumber\\
    B_{1,1;1}(\{x^{(i,j)}\})&=&x^{(1,1)}\,,\nonumber\\
    B_{1,1;2}(\{x^{(i,j)}\})&=&x^{(1,0)}\,x^{(0,1)}\,,\nonumber\\
    B_{3,0;1}(\{x^{(i,j)}\})&=&x^{(3,0)}\,,\nonumber\\
    B_{3,0;2}(\{x^{(i,j)}\})&=&3\,x^{(1,0)}\,x^{(2,0)}\,,\nonumber\\
    B_{3,0;3}(\{x^{(i,j)}\})&=&(x^{(1,0)})^3\,,\nonumber\\
    B_{2,1;1}(\{x^{(i,j)}\})&=&x^{(2,1)}\,,\nonumber\\
    B_{2,1;2}(\{x^{(i,j)}\})&=&2\,x^{(1,0)}\,x^{(1,1)}+x^{(0,1)}\,x^{(2,0)},\nonumber\\
    B_{2,1;3}(\{x^{(i,j)}\})&=&x^{(0,1)}\,(x^{(1,0)})^2.\nonumber\\
    %\end{eqnarray}
    %\begin{eqnarray}
    B_{1,2;1}(\{x^{(i,j)}\})&=&x^{(1,2)}\,,\nonumber\\
    B_{1,2;2}(\{x^{(i,j)}\})&=&2\,x^{(0,1)}\,x^{(1,1)}+x^{(1,0)}\,x^{(0,2)},\nonumber\\
    B_{1,2;3}(\{x^{(i,j)}\})&=&x^{(1,0)}\,(x^{(0,1)})^2.\nonumber\\
    B_{0,3;1}(\{x^{(i,j)}\})&=&x^{(0,3)}\,,\nonumber\\
    B_{0,3;2}(\{x^{(i,j)}\})&=&3\,x^{(0,1)}\,x^{(0,2)}\,,\nonumber\\
    B_{0,3;3}(\{x^{(i,j)}\})&=&(x^{(0,1)})^3\,,\nonumber
\end{eqnarray}

A combinatorial interpretation of the Bell polynomial $B_{n,m;k}({x^{\{i,j\}}})$ goes as follows. Consider a collection of $n$ blue beads and $m$ red ones. How can these be split into $k$ groups is encoded in the Bell polynomials. A group consisting of $i$ blue beads and $j$ red ones is denoted by $x^{\{i,j\}}$. For instance, a system consisting of two blue beads and one red one can be split into two groups in three ways. Two with one blue and one red bead in one group and the remaining blue bead in the other group and one with the two blue beads in one group and the red one in the other. This case corresponds to the Bell polynomial $B_{2,1;2}(\{x^{(i,j)}\})$ in (\ref{eq:app-7}).

The generalized Fa\`{a} di Bruno formula for computing the derivatives of a composite function of two variables, $f(g(z_1,z_2))$, needed for the calculation of the factorials cumulants is then,
\begin{eqnarray}\label{eq:faa-di-bruno-mult1}
    \frac{\partial^{n_1}}{\partial z_1^{n_1}}\,\frac{\partial^{n_2}}{\partial z_2^{n_2}}f(g(z_1,z_2))=\sum_{k=1}^{n_1+n_2}f^{(k)}\,B_{n_1,n_2;k}(\{g^{(i,j)}\}).
\end{eqnarray}
The first few terms are
\begin{eqnarray}
    \frac{\partial}{\partial z_1}\,f(g(z_1,z_2))&=&f^{(1)}g^{(1,0)}\,,\nonumber\\
    \frac{\partial}{\partial z_2}\,f(g(z_1,z_2))&=&f^{(1)}g^{(0,1)}\,,\nonumber\\
    \frac{\partial^2}{\partial z_1^2}\,f(g(z_1,z_2))&=&f^{(1)}g^{(2,0)}+f^{(2)}\big(g^{(1,0)}\big)^2\,,\\
    \frac{\partial^2}{\partial z_2^2}\,f(g(z_1,z_2))&=&f^{(1)}g^{(0,2)}+f^{(2)}\big(g^{(0,1)}\big)^2\,,\nonumber\\
    \frac{\partial}{\partial z_1}\frac{\partial}{\partial z_2}\,f(g(z_1,z_2))&=&f^{(1)}g^{(1,1)}+f^{(2)}\,g^{(1,0)}\,g^{(0,1)}\,.\nonumber
\end{eqnarray}

We also need the Bell polynomials for computing the derivatives of a function of the form $f(g_1(z),g_2(z))$. The generating function is:
\begin{eqnarray}
    \Phi(t,u_1,u_2)&=&\exp\left(\sum_{j=1}^\infty \big[u_1*x_1^{(j)}+u_2*x_2^{(j)}\big]\frac{t^{j}}{j!}\right)\\
    &=&\sum_{\substack{n,k_1,k_2\\ n\geq k_1+k_2\geq 0}}B_{n;k_1,k_2}\big(\{\vec{x}^{(i)}\}\big)\,\frac{t^{n}}{n!}\,(u_1^{k_1}+u_2^{k_2}).\nonumber
\end{eqnarray}
and the corresponding generalized Bell polynomials are given by,
\begin{equation}
    B_{n;k_1,k_2}(\{\vec{x}^{(i)}\})=\frac{1}{k_1!\,k_2!}
    \frac{\partial^{n}}{\partial t^{n}}\frac{\partial^{k_1}}{\partial u_1^{k_1}}\frac{\partial^{k_2}}{\partial u_2^{k_2}}\,\Phi(t_1,t_2,u)|_{t_1=t_2=u=0}.
\end{equation}
We note that the multivariate Bell polynomials of this type can be constructed from the standard (univariate) Bell polynomials,~\cite{Riordan:1946}
\begin{eqnarray}\label{eq:bell-construct}
    &&B_{n;k_1,k_2}(x_1^{(1)},x_1^{(2)},\dots;x_2^{(1)},x_2^{(2)},\dots)\\
    &&=\sum_{i=0}^n\binom{n}{i}\,B_{i,k_1}(x_1^{(1)},x_1^{(2)},\dots)\,B_{n-i,k2}(x_2^{(1)},x_2^{(2)},\dots).\nonumber
\end{eqnarray}

The first few Bell polynomials are
\begin{eqnarray}
    B_{1;1,0}(\{\vec{x}^{(i)}\})&=&x_1^{(1)},\nonumber\\
    B_{1;0,1}(\{\vec{x}^{(i)}\})&=&x_2^{(1)},\nonumber\\
    B_{2;1,0}(\{\vec{x}^{(i)}\})&=&x_1^{(2)},\nonumber\\
    B_{2;0,1}(\{\vec{x}^{(i)}\})&=&x_2^{(2)},\\
    B_{2;2,0}(\{\vec{x}^{(i)}\})&=&(x_1^{(1)})^2,\nonumber\\
    B_{2;0,2}(\{\vec{x}^{(i)}\})&=&(x_2^{(1)})^2,\nonumber\\
    B_{2;1,1}(\{\vec{x}^{(i)}\})&=&2\,x_1^{(1)}\,x_2^{(1)},\nonumber
\end{eqnarray}
and the corresponding Fa\`{a} di Bruno formula is~\cite{Riordan:1946}
\begin{eqnarray}\label{eq:faa-di-bruno-mult2}
    \frac{\partial^{n}}{\partial z^{n}} f(g_1(z),g_2(z))=\sum_{\substack{k_1,k_2\\ k_1+k_2\geq 0}}^{n}f^{(k_1,k_2)}\,B_{n;k_1,k_2}(\{g_1^{(i)};g_2^{(i)}\}).
\end{eqnarray}
Here the first few terms are given by
\begin{eqnarray}
    \frac{\partial}{\partial z}f(g_1(z),g_2(z))&=&f^{(1,0)}g_1^{(1)}+f^{(0,1)}g_2^{(1)}\,,\nonumber\\
    \frac{\partial^2}{\partial z^2}f(g_1(z),g_2(z))&=&f^{(1,0)}g_1^{(2)}+f^{(0,1)}g_2^{(2)}+f^{(2,0)}\big(g_1^{(1)}\big)^2\\
    &+&f^{(0,2)}\big(g_2^{(1)}\big)^2+2\,f^{(1,1)}\,g_1^{(1)}\,g_2^{(1)}\,.\nonumber
\end{eqnarray}

\section{Generalized Stirling numbers}\label{sec:app-b}

Consider multivariate Bell polynomials of the type (\ref{eq:gen-Bell-21}), with the derivatives $x^{(i,j)}$ given by $g^{(i,j)}$ in
(\ref{eq:gc-derivatives}). In this case, the generating function (\ref{eq:gen-function-multiv-Bell-21}) reduces to 
\begin{equation}
 \Phi(t_1,t_2,u)=\big[(1+\alpha_B\, t_1)\,(1+\alpha_{\bar B}\,t_2)\big]^{u/2},   
\end{equation}
and the corresponding Bell polynomials are
\begin{equation}
    B_{n,m;k}(\{g_3^{(i,j)}\})=\frac{(\alpha_B)^n\,(\alpha_{\bar B})^m}{2^k}\,\,s(n,m;k),
\end{equation}
where
\begin{equation}
    s(n,m;k)=\frac{1}{k!}\,\frac{\partial^n}{\partial z^n}\,\frac{\partial^m}{\partial w^m}\,\frac{\partial^k}{\partial u^k}\,\Psi(z,w,u)|_{z=w=u=0},
\end{equation}
are generalized Stirling numbers and 
\begin{equation}
    \Psi(z,w,u)=\big[(1+z)\,(1+w)\big]^u
\end{equation}
is the corresponding generating function.  

The generating function for the regular Stirling numbers of the first kind
\begin{equation}
    s(n,k)=\frac{1}{k!}\,\frac{\partial^n}{\partial z^n}\,\frac{\partial^k}{\partial u^k}\,\psi(z,u)|_{z=u=0}
\end{equation}
is of the form~\cite{Comtet:1974}
\begin{equation}
    \psi(z,u)=\big[1+z\big]^u.
\end{equation}
It follows that 
\begin{equation}\label{eq:gen-reg-stirling}
 s(n,0;k)=s(0,n;k)=s(n,k),   
\end{equation}
that $s(n,m;0)=0$, except for $n=m=0$, and that $s(n,m;k)=0$ for $k>n+m$, and that 
\begin{equation}
    s(n,m;k)=\sum_{l=0}^{k}\,s(n,k-l)\,s(m,l).
\end{equation}

Moreover, the generalized Stirling numbers satisfy the recurrence relations
\begin{eqnarray}\label{eq:gen-stirling-recurrence-1}
    s(n,m;k)&=&s(n-1,m;k-1)-(n-1)\,s(n-1,m;k),\\
    s(n,m;k)&=&s(n,m-1;k-1)-(m-1)\,s(n,m-1;k),\label{eq:gen-stirling-recurrence-2}
   % \nonumber
\end{eqnarray}
in close analogy to the one obeyed by the regular Stirling  numbers of the first kind~\cite{Comtet:1974}. The relation (\ref{eq:gen-stirling-recurrence-1}) holds for $n,k\geq 1$ and (\ref{eq:gen-stirling-recurrence-2}) for $m,k\geq 1$.
The two relations (\ref{eq:gen-stirling-recurrence-1}) and (\ref{eq:gen-stirling-recurrence-2}) combined yield a recurrence relation at fixed $k$,
\begin{equation}\label{eq:gen-stirling-recurrence-3}
    s(n,m;k)=s(n+1,m-1;k)+(n-m+1)\,s(n,m-1;k).
\end{equation}
The recurrence relation for factorial cumulants,
\begin{equation}
    F^{(n+1,m)}=\frac{\alpha_B}{\alpha_{\bar B}}\,F^{(n,m+1)}-(n-m)\,\alpha_B\,F^{(n,m)},
\end{equation}
which was empirically deduced in~\cite{Barej:2020ymr}, follows from (\ref{eq:gen-stirling-recurrence-3}).

The generalized Stirling numbers also have a combinatorial interpretation. The absolute value, $|s(n,m;k)|$, equals the number of permutations of $n$ blue beads and $m$ red beads in k disjoint cycles, where each cycle consists of only blue or only red beads. The phase of $s(n,m;k)$ reproduces the phase stemming from the derivatives of $g_3(x,\bar{x})$, given in (\ref{eq:gc-derivatives}).

\section{General relations between cumulants and factorial cumulants}\label{sect:App-c}

In this appendix we derive general relations between cumulants and factorial cumulants, which are independent of the assumed partition function. We start by considering the fluctuations of $N^{(tot)}=N_B+N_{\bar{B}}$. The cumulants $C_n$ and factorial cumulants $F_n$ are obtained by differentiating the corresponding generating functions,
\begin{equation}\label{eq:cumulant-gen-func-Ntot}
    C_n=\frac{d^n G(t)}{d\,t^n}\mid_{t=0}.
\end{equation}
and
\begin{equation}\label{eq:fac-cumulants-gen-func-Ntot}
    F_n=\frac{d^n H(x)}{d\,x^n}\mid_{x=1},
\end{equation}
which are closely related through~\cite{Kitazawa:2017ljq}
\begin{equation}\label{eq:cum-fac-cum-gen-func}
    G(t)=H(e^t),\qquad H(x)=G(\ln(x)).
\end{equation}
By applying the Fa\`{a} di Bruno formula to composite functions of the form $f(\ln x)$ and $f(e^t)$ one finds the following general relations between cumulants and factorial cumulants~\cite{Comtet:1974},
\begin{equation}\label{eq:Cn_to_Fn-relations}
    F_n=\sum_{k=1}^k\,s(n,k)\,C_k,\qquad C_n=\sum_{k=1}^n\,S(n,k)\,F_k,
\end{equation}
where $s(n,k)$ and $S(n,k)$ are Stirling numbers of the first and second kind, respectively. The Stirling numbers satisfy the orthogonality relations~\cite{Comtet:1974}
\begin{eqnarray}\label{eq:orth-Stirling-1}
    \sum_{k=0}^n\,S(n,k)\,s(k,m)&=&\delta_{n,m},\\
    \sum_{k=0}^n\,s(n,k)\,S(k,m)&=&\delta_{n,m}.\label{eq:orth-Stirling-2}
\end{eqnarray}

Analogously, one finds the relation between the cumulants and factorial cumulants of a subsystem by using the multivariant Fa\'{a} di Bruno formula (\ref{eq:faa-di-bruno-mult1}) for a function of the form $G(e^t,e^s)$,
\begin{equation}\label{eq:cumulant-fac-cumulant-relation-subsystem}
    C^{(n,m)}=\sum_{k_1=0}^n\,\sum_{k_2=0}^m\,F^{(k1,k2)}\,S(n,k_1)\,S(m,k_2).
\end{equation}
For completeness we note that the relation (\ref{eq:cumulant-fac-cumulant-relation-subsystem}) can be inverted, using the orthogonality property of the Stirling numbers, (\ref{eq:orth-Stirling-2}),
\begin{equation}\label{eq:fac-cumulant-cumulant-relation-subsystem}
    F^{(p,q)}=\sum_{n=0}^p\,\sum_{m=0}^q\,C^{(n,m)}\,s(p,n)\,s(q,m).
\end{equation}

Utilizing the fact that cumulants of the net-baryon number correspond to cumulants  of  $\delta N_B - \delta N_{\bar B}$, one finds, using the binomial theorem, the general relation
\begin{equation}\label{eq:kappa-Cmn}
    \kappa_n=\sum_{i=0}^n\,\binom{n}{i}\,(-1)^{n-i}\,C^{(i,n-i)},
\end{equation}
where $\binom{n}{i}$ is a binomial coefficient. Similarly, one obtains the cumulants $\kappa^s_n$ of $\delta (N_B+N_{\bar B})$ in the subsystem using
\begin{equation}
    \kappa^s_n=\sum_{i=0}^n\,\binom{n}{i}\,C^{(i,n-i)}.
\end{equation}
In the limit $\alpha_B,\alpha_{\bar B} \to 1$, $\kappa_n\to 0$, while $\kappa^s_n \to C_n$. 

Finally, by inserting (\ref{eq:cumulant-fac-cumulant-relation-subsystem}) in (\ref{eq:kappa-Cmn}) one finds a general relation between the factorial cumulants and the net-baryon-number cumulants~\cite{Luo:2014rea},
\begin{equation}\label{eq:cumulant-factorial-cumulant-relation-appendix}
    \kappa_n=\sum_{\substack{k_1,k_2\\ k_1+k_2\geq 0}}^{n}F^{(k_1,k_2)}\,\sum_{i=0}^n\binom{n}{i}\,(-1)^{n-i}\,S(i,k_1)\,S(n-i,k_2).
\end{equation}

%\begin{thebibliography}{99}
%\bibliography{references}
%

\end{document}